\documentclass[a4paper,12pt]{article} 
\usepackage{amsfonts,slashed}
\usepackage{url}
\usepackage{latexsym}
\usepackage{amsmath}
\usepackage{amssymb}
\usepackage{indentfirst}
\usepackage{graphicx}
\usepackage{epsfig}
\usepackage{amsthm}
\usepackage{amsfonts}
\usepackage{indentfirst}
\usepackage[colorlinks]{hyperref}
\usepackage{cite}
\usepackage{cancel}
\usepackage{tikz}

\begin{document}

\begin{titlepage}

\date{}

\title{\bf{Generalized AdS-Lorentz deformed supergravity on a manifold with boundary}}
\author{Alessandro Banaudi$^{1,2}$\thanks{alessandro.banaudi@mi.infn.it}, Lucrezia Ravera$^{1}$\thanks{lucrezia.ravera@mi.infn.it} \\
{\small $^{1}$\textit{INFN, Sezione di Milano, Via Celoria 16, I-20133 Milano, Italy}}\\
{\small $^{2}$\textit{Dipartimento di Fisica, Universit\`{a} di Milano, Via Celoria 16, 20133 Milano, Italy}}}
\clearpage\maketitle
\thispagestyle{empty}

\begin{abstract}

The purpose of this paper is to explore the supersymmetry invariance of a particular supergravity theory, which we refer to as $D=4$ generalized AdS-Lorentz deformed supergravity, in the presence of a non-trivial boundary. In particular, we show that the so-called generalized minimal AdS-Lorentz superalgebra can be interpreted as a peculiar torsion deformation of $\mathfrak{osp}(4\vert 1)$, and we present the construction of a bulk Lagrangian based on the aforementioned generalized AdS-Lorentz superalgebra. 
In the presence of a non-trivial boundary of space-time, that is when the boundary is not thought of as set at infinity, the fields do not asymptotically vanish, and this has some consequences on the invariances of the theory, in particular on supersymmetry invariance. In this work, we adopt the so-called rheonomic (geometric) approach in superspace and show that a supersymmetric extension of a Gauss-Bonnet like term is required in order to restore the supersymmetry invariance of the theory. The action we end up with can be recast as a MacDowell-Mansouri type action, namely as a sum of quadratic terms in the generalized AdS-Lorentz covariant super field-strengths.

\end{abstract}

\vspace{2cm}

\noindent

\end{titlepage}

\section{Introduction}\label{introduction}

Gravity and supergravity theories in diverse dimensions in the presence of a boundary have been studied in different contexts in the last forty years (see, for example, \cite{York:1972sj, Gibbons:1976ue, Brown:1992br, Horava:1996ma}). 

A particularly relevant field in which they find application is the so-called AdS/CFT duality (see the first works \cite{Maldacena:1997re, Gubser:1998bc, Witten:1998qj, Aharony:1999ti, DHoker:2002nbb} on this topic and references therein). 
In the supergravity limit (i.e. low-energy limit) of string theory, this duality implies a one-to-one correspondence between quantum operators in the CFT on the boundary and
the fields of the supergravity theory in the bulk. In AdS/CFT, the action functional is required to be supplemented with proper boundary conditions for the supergravity fields, the latter acting as sources for the CFT operators. The divergences of the bulk metric near the boundary can be eliminated through the so-called holographic
renormalization (see, for instance, \cite{Skenderis:2002wp} and references therein), with the inclusion of appropriate counterterms at the boundary.

In relevant works such as \cite{Aros:1999id, Aros:1999kt, Mora:2004kb, Olea:2005gb, Jatkar:2014npa}, the inclusion of boundary terms and counterterms to AdS gravity was studied, and, on the other hand, many authors \cite{vanNieuwenhuizen:2005kg, Belyaev:2005rt, Belyaev:2007bg, Belyaev:2008ex, Grumiller:2009dx, Belyaev:2010as} considered it in the context of supergravity theories, by adopting different approaches. 
The results of these works pointed out to the conclusion that, in order to restore all the invariances of a
(super)gravity Lagrangian with cosmological constant on a manifold with a non-trivial boundary (that is when the boundary is not thought as set at infinity), one needs to add topological (i.e. boundary) contributions to the theory, also providing the counterterms
necessary for regularizing the action and the conserved charges.

More recently, in \cite{Andrianopoli:2014aqa} the authors constructed the $\mathcal{N} = 1$ and $\mathcal{N} = 2$, $D = 4$ supergravity theories with negative cosmological constant in the presence of a non-trivial boundary in a geometric framework (extending to superspace the geometric approach of \cite{Aros:1999id, Aros:1999kt, Mora:2004kb, Olea:2005gb, Jatkar:2014npa}): Precisely, they generalized the so-called rheonomic (geometric) approach to supergravity \cite{Castellani:1991eu} (see also \cite{Castellani:2018zey, Ravera:2018zvm} for recent reviews of this framework) in the presence of a non-trivial boundary and they added proper boundary terms to the Lagrangian in order to restore the supersymmetry invariance of the theory. In particular, the authors found that the supersymmetry invariance of the full Lagrangian (understood as bulk plus boundary contributions) is recovered with the introduction of a supersymmetric extension of the Gauss-Bonnet term. The final Lagrangian is written down as a sum of quadratic terms in $OSp(\mathcal{N} \vert 4)$-covariant super field-strengths, reproducing a MacDowell-Mansouri type action \cite{MacDowell:1977jt}.

Lately, in \cite{Ipinza:2016con} the authors explored the supersymmetry invariance of a particular supergravity theory in the
presence of a non-trivial boundary, following the prescription of \cite{Andrianopoli:2014aqa}. Specifically, they presented the explicit construction of a geometric bulk Lagrangian based on an enlarged superalgebra, known as AdS-Lorentz superalgebra, showing that, also in this case, the supersymmetric extension of a Gauss-Bonnet like term is
required to restore the supersymmetry invariance of the complete theory. In analogy to the result of \cite{Andrianopoli:2014aqa}, they obtained that the full action can be finally written as a MacDowell-Mansouri type action.

Driven by the results of \cite{Andrianopoli:2014aqa, Ipinza:2016con} (see also \cite{Ravera:2018zvm}), in this work we explore the supersymmetry invariance of a supergravity theory we will refer to as $D=4$ generalized AdS-Lorentz deformed supergravity, in the rheonomic approach in the presence of a non-trivial boundary. In particular, we present the construction of a geometric bulk Lagrangian based on the generalized minimal AdS-Lorentz superalgebra introduced in \cite{Concha:2015tla}, which is larger than $\mathfrak{osp}(4\vert 1)$ and, as we will explicitly show in the sequel, can be seen as a peculiar torsion deformation of $\mathfrak{osp}(4\vert 1)$. Then, we study the supersymmetry invariance of the Lagrangian in the presence of a non-trivial space-time boundary.

The present paper is organized as follows: In Section \ref{AdS-Lpres}, we recall some aspects of AdS-Lorentz superalgebras, showing that they can be seen as particular torsion deformations of the AdS superalgebra $\mathfrak{osp}(4\vert 1)$. To this aim, we write and analyze their dual Maurer-Cartan formulation. In Section \ref{rheon}, we present the explicit geometric construction of the bulk Lagrangian in terms of the generalized AdS-Lorentz supercurvatures, in which a scale parameter $e$ appears. Then, we show that the same Lagrangian can be rewritten in terms of Lorentz-type curvatures for which $e \rightarrow 0$. The whole procedure can be viewed as an alternative way to introduce a generalized cosmological constant in the theory.
Subsequently, in Section \ref{bdysec}, we study the supersymmetry invariance of the Lagrangian in the presence of a non-trivial boundary of space-time. In particular, we show that, in order to restore the supersymmetry invariance
of the full Lagrangian, a supersymmetric Gauss-Bonnet like term is necessary. The action obtained in this way can be finally recast in a suggestive form as a sum of quadratic terms in generalized AdS-Lorentz covariant super field-strengths, that is as a MacDowell-Mansouri type action \cite{MacDowell:1977jt}. 
Section \ref{conclusions} contains our conclusions and possible future developments, while in Appendix \ref{appa} we collect some useful formulas in $D=4$ space-time dimensions.

\section{AdS-Lorentz superalgebras and some of their features}\label{AdS-Lpres}

In this section, we recall some features of the so-called AdS-Lorentz superalgebra and of its minimal generalization. We also write the dual Maurer-Cartan form of the aforementioned superalgebras and show that they can be seen as peculiar torsion deformations of $\mathfrak{osp}(4 \vert 1)$.

The AdS-Lorentz (super)algebra was obtained as a deformation of the Maxwell (super)symmetries \cite{Soroka:2006aj, Durka:2011nf} and it can be alternatively derived through a particular expansion process, called $S$-expansion \cite{Izaurieta:2006zz},\footnote{The $S$-expansion method \cite{Izaurieta:2006zz} is based on combining the multiplication law of a semigroup $S$ with the structure constants of a Lie (super)algebra $\mathfrak{g}$, in such a way to end up with a new, larger, Lie (super)algebra $\mathfrak{g}_S = S \times \mathfrak{g}$, that is called the $S$-expanded (super)algebra (see also \cite{Ipinza:2016bfc} for an analytic method for performing $S$-expansion).} of the AdS (super)algebra \cite{Concha:2015tla, Diaz:2012zza, Fierro:2014lka, Concha:2016hbt}. 
When the AdS-Lorentz algebra is considered, it is possible to introduce a generalized cosmological constant term in a Born-Infeld like gravity action \cite{Salgado:2014qqa, Concha:2016tms, Concha:2017nca}; analogously, the AdS-Lorentz superalgebra and its minimal generalization allow to introduce a generalized supersymmetric cosmological constant term in $\mathcal{N}=1$, $D=4$ supergravity \cite{Concha:2015tla}.

The AdS-Lorentz superalgebra is generated by the set $\left\{
J_{ab},P_{a},Z_{ab},Q_{\alpha }\right\} $ ($a=0,1,2,3$ and $\alpha=1,2,3, 4$ in $D=4$), it is semisimple, and its (anti)commutation relations read
\begin{equation}\label{adslalg}
\begin{split}
\left[ J_{ab},J_{cd}\right] & =\eta _{bc}J_{ad}-\eta _{ac}J_{bd}-\eta
_{bd}J_{ac}+\eta _{ad}J_{bc}, \\
\left[ J_{ab},Z_{cd}\right] & =\eta _{bc}Z_{ad}-\eta _{ac}Z_{bd}-\eta
_{bd}Z_{ac}+\eta _{ad}Z_{bc}, \\
\left[ Z_{ab},Z_{cd}\right] & =\eta _{bc}Z_{ad}-\eta _{ac}Z_{bd}-\eta
_{bd}Z_{ac}+\eta _{ad}Z_{bc},  \\
\left[ J_{ab},P_{c}\right] & =\eta _{bc}P_{a}-\eta _{ac}P_{b}, \quad \left[ P_{a},P_{b}\right] = Z_{ab}, \quad \left[ Z_{ab},P_{c}\right]  =\eta _{bc}P_{a}-\eta _{ac}P_{b}, \\
\left[ J_{ab},Q_{\alpha }\right] & =-\frac{1}{2}\left( \gamma _{ab}Q\right)_{\alpha }, \quad \left[ P_{a},Q_{\alpha }\right]  =-\frac{1}{2} \left( \gamma _{a}Q\right) _{\alpha }, \quad
\left[ Z_{ab},Q_{\alpha }\right]  =-\frac{1}{2}\left( \gamma _{ab}Q\right) _{\alpha }, \\
\left\{ Q_{\alpha },Q_{\beta }\right\} & =-\frac{1}{2}\left[ \left( \gamma ^{ab}C\right) _{\alpha \beta }Z_{ab} -2\left( \gamma ^{a}C\right) _{\alpha \beta }P_{a}\right], 
\end{split}
\end{equation}
where $C$ is the charge conjugation matrix, $\gamma _{a}$ and $\gamma_{ab}$ are gamma matrices in four dimensions, $J_{ab}$ and $P_a$ are the Lorentz and translations generators, respectively, $Q_\alpha$ is the supersymmetry charge, and $Z_{ab}$ are non-abelian Lorentz-like generators. 

The generators $\lbrace P_a, Z_{ab}, Q_\alpha \rbrace$ span a non-abelian ideal of the AdS-Lorentz superalgebra \eqref{adslalg}.
Let us also observe that the Lorentz-type algebra $\mathcal{L}=\left\{ J_{ab},Z_{ab}\right\} $ is a subalgebra of \eqref{adslalg}. This subalgebra and its extensions to higher dimensions have been useful to derive General Relativity from Born-Infeld gravity theories \cite{Concha:2013uhq, Concha:2014vka, Concha:2014zsa}.

The minimal generalization of the AdS-Lorentz superalgebra \eqref{adslalg} contains one more spinor charge and it can be found in \cite{Concha:2015tla}, where it was obtained through the so-called $S$-expansion procedure from $\mathfrak{osp}(4 \vert 1)$.\footnote{In the sequel, we will refer to this minimal generalization of the AdS-Lorentz superalgebra as the generalized minimal AdS-Lorentz superalgebra or, for simplicity, just as the generalized AdS-Lorentz superalgebra.}
Let us mention that an In\"{o}n\"{u}-Wigner contraction of the generalized minimal AdS-Lorentz superalgebra leads to a generalization of the minimal Maxwell superalgebra introduced in \cite{Bonanos:2009wy}. The Maxwell algebra \cite{Bacry:1970ye, Schrader:1972zd, Bonanos:2008kr, Bonanos:2010fw, Gomis:2009vm, Bonanos:2008ez, Gibbons:2009me} (see also the more recent paper \cite{Gomis:2017cmt}) is a non-central extension of the Poincar\'{e} algebra\footnote{In fact, the Maxwell algebra is obtained from the Poincar\'{e} algebra by replacing the commutator $[P_a,P_b]=0$ of the latter with $[P_a,P_b] = Z_{ab}$, where $Z_{ab} = -Z_{ba}$ are abelian generators commuting with translations and behaving like tensors with
respect to Lorentz transformations.} and it describes the symmetries of systems evolving in flat Minkowski space filled in by a constant electromagnetic background. The minimal supersymmetric extension of the Maxwell algebra involves an extra spinor charge (besides the spinor charge $Q_\alpha$ of the super-Poincar\'{e} algebra) \cite{Bonanos:2009wy}.

The generalized minimal AdS-Lorentz superalgebra is generated by the set \\ $\left\{
J_{ab},P_{a},\tilde{Z}_a,\tilde{Z}_{ab},Z_{ab},Q_{\alpha }, \Sigma_{\alpha }\right\} $ ($a=0,1,2,3$ and $\alpha=1,2,3, 4$ in $D=4$) \\ and its (anti)commutation relations read as follows:
\begin{equation}\label{genadslalg}
\begin{split}
\left[ J_{ab},J_{cd}\right] & =\eta _{bc}J_{ad}-\eta _{ac}J_{bd}-\eta
_{bd}J_{ac}+\eta _{ad}J_{bc}, \\
\left[ Z_{ab},Z_{cd}\right] & =\eta _{bc}Z_{ad}-\eta _{ac}Z_{bd}-\eta
_{bd}Z_{ac}+\eta _{ad}Z_{bc},  \\
\left[ J_{ab},Z_{cd}\right] & =\eta _{bc}Z_{ad}-\eta _{ac}Z_{bd}-\eta
_{bd}Z_{ac}+\eta _{ad}Z_{bc}, \\
\left[ J_{ab},\tilde{Z}_{cd}\right] & =\eta _{bc}\tilde{Z}_{ad}-\eta _{ac}\tilde{Z}_{bd}-\eta _{bd}\tilde{Z}_{ac}+\eta _{ad}\tilde{Z}_{bc}, \\
\left[ \tilde{Z}_{ab},\tilde{Z}_{cd}\right] & =\eta _{bc}Z_{ad}-\eta _{ac}Z_{bd}-\eta _{bd}Z_{ac}+\eta _{ad}Z_{bc},  \\
\left[ \tilde{Z}_{ab}, Z_{cd}\right] & =\eta _{bc}\tilde{Z}_{ad}-\eta _{ac}\tilde{Z}_{bd}-\eta _{bd}\tilde{Z}_{ac}+\eta _{ad}\tilde{Z}_{bc},  \\
\left[ J_{ab},P_{c}\right] & =\eta _{bc}P_{a}-\eta _{ac}P_{b}, \quad \left[ Z_{ab},P_{c}\right] =\eta _{bc}P_{a}-\eta _{ac}P_{b}, \\
\left[ \tilde{Z}_{ab},P_{c}\right] & =\eta _{bc}\tilde{Z}_{a}-\eta _{ac}\tilde{Z}_{b}, \quad
\left[ J_{ab},\tilde{Z}_{c}\right] =\eta _{bc}\tilde{Z}_{a}-\eta _{ac}\tilde{Z}_{b}, \\
\left[ \tilde{Z}_{ab},\tilde{Z}_{c}\right] & =\eta _{bc}P_{a}-\eta _{ac}P_{b}, \quad 
\left[ Z_{ab},\tilde{Z}_{c}\right]  =\eta _{bc}\tilde{Z}_{a}-\eta _{ac}\tilde{Z}_{b}, \\
\left[ P_{a},P_{b}\right] &= Z_{ab}, \quad \left[\tilde{Z}_{a},P_{b}\right]  = \tilde{Z}_{ab}, \quad \left[ \tilde{Z}_{a},\tilde{Z}_{b}\right] = Z_{ab}, \\
\left[ J_{ab},Q_{\alpha }\right] & =-\frac{1}{2}\left( \gamma _{ab}Q\right)_{\alpha }, \quad
\left[ P_{a},Q_{\alpha }\right] =-\frac{1}{2} \left( \gamma _{a}\Sigma\right) _{\alpha }, \\
\left[ \tilde{Z}_{ab},Q_{\alpha }\right] & =-\frac{1}{2}\left( \gamma _{ab}\Sigma\right)_{\alpha }, \quad \left[ \tilde{Z}_{a},Q_{\alpha }\right] =-\frac{1}{2} \left( \gamma _{a}Q\right) _{\alpha }, \\
\left[ Z_{ab},Q_{\alpha }\right] & =-\frac{1}{2}\left( \gamma _{ab}Q\right)_{\alpha }, \quad
\left[ P_{a},\Sigma_{\alpha }\right] =-\frac{1}{2} \left( \gamma _{a}Q\right) _{\alpha }, \\
\left[ J_{ab},\Sigma_{\alpha }\right] & =-\frac{1}{2}\left( \gamma _{ab}\Sigma\right) _{\alpha }, \quad
\left[ \tilde{Z}_{a},\Sigma_{\alpha }\right]  =-\frac{1}{2} \left( \gamma _{a}\Sigma\right) _{\alpha }, \\
\left[ \tilde{Z}_{ab},\Sigma_{\alpha }\right]  &=-\frac{1}{2}\left( \gamma _{ab}Q\right)_{\alpha }, \quad
\left[ Z_{ab},\Sigma_{\alpha }\right]  =-\frac{1}{2}\left( \gamma _{ab}\Sigma\right)_{\alpha }, \\
\left\{ Q_{\alpha },Q_{\beta }\right\} & =-\frac{1}{2}\left[ \left( \gamma ^{ab}C\right) _{\alpha \beta }\tilde{Z}_{ab}-2\left( \gamma ^{a}C\right) _{\alpha \beta }P_{a}\right] , \\
\left\{ Q_{\alpha },\Sigma_{\beta }\right\} & =-\frac{1}{2}\left[ \left( \gamma ^{ab}C\right) _{\alpha \beta }Z_{ab}-2\left( \gamma ^{a}C\right) _{\alpha \beta }\tilde{Z}_{a}\right] , \\
\left\{ \Sigma_{\alpha },\Sigma_{\beta }\right\} & =-\frac{1}{2}\left[ \left( \gamma
^{ab}C\right) _{\alpha \beta }\tilde{Z}_{ab}-2\left( \gamma ^{a}C\right) _{\alpha
\beta }P_{a}\right].
\end{split}
\end{equation}
As we can see above, a new Majorana spinor charge appears. The introduction of a second spinorial generator can also be found, for example, in \cite{DAuria:1982uck, Andrianopoli:2016osu, Penafiel:2017wfr, Andrianopoli:2017itj, Ravera:2018vra} (see also \cite{Ravera:2018zvm}) and \cite{Green:1989nn} in the supergravity and superstring contexts, respectively). 

Notice that by setting $\tilde{Z}_a \rightarrow 0$ the Jacobi identities of \eqref{genadslalg} are still fulfilled.
Let us also observe, as it was already pointed out in \cite{Concha:2015tla}, that the generalized AdS-Lorentz algebra $\lbrace J_{ab}, P_a, \tilde{Z}_a, \tilde{Z}_{ab},Z_{ab} \rbrace$ and the algebra $\lbrace J_{ab}, P_a, Z_{ab} \rbrace$ are bosonic subalgebras of \eqref{genadslalg}. 
Furthermore, an In\"{o}n\"{u}-Wigner contraction of \eqref{genadslalg} provides the so-called minimal Maxwell superalgebra $s\mathcal{M}_4$ of \cite{Concha:2014tca} (namely a minimal generalization of the Maxwell superalgebra).

\subsection{The AdS-Lorentz superalgebra as a torsion deformation of $\mathfrak{osp}(4\vert 1)$}


Before moving to the analysis of the supersymmetry invariance of a deformed $D=4$ supergravity theory based on the generalized minimal AdS-Lorentz superalgebra \eqref{genadslalg} in the presence of a non-trivial boundary, we clarify in the following the relations between the AdS-Lorentz superalgebras \eqref{adslalg} and \eqref{genadslalg} and $\mathfrak{osp}(4\vert 1)$.

Let us first consider the AdS-Lorentz superalgebra, given in \eqref{adslalg}. 

We introduce the set of $1$-forms $\left\{\omega^{ab},V^{a},k^{ab},\psi^{\alpha }\right\}$, that are $1$-form fields respectively dual to the generators $\left\{J_{ab},P_{a},Z_{ab},Q_{\alpha }\right\}$,\footnote{In the sequel, for simplifying our notation, we will neglect the spinor index $\alpha$.} that is
\begin{equation}
\omega^{ab} (J_{cd}) = \delta^{ab}_{cd}, \quad V^a (P_b) = \delta^a_b, \quad k^{ab} (Z_{cd}) = \delta^{ab}_{cd}, \quad \psi (Q) = \mathfrak{1}.
\end{equation}
Observe, in particular, that the presence of the bosonic generator $Z_{ab}$ implies the introduction of its dual $1$-form field $k^{ab}$.

The aforementioned $1$-form fields obey the following Maurer-Cartan equations:
\begin{subequations}\label{mcadsl}
\begin{align}
& d\omega ^{ab}+\omega ^{ac} \wedge \omega _c^{\;b} =0, \label{mcadslomega} \\
& D_\omega V^{a}+k _{\;b}^{a} \wedge V^{b}-\frac{1}{2}\bar{\psi} \wedge \gamma^{a} \psi =0, \label{mcadslv} \\
& D_\omega k^{ab}+k _{\;c}^{a}\wedge k^{cb}+4 e^{2}\; V^{a} \wedge V^{b}+ e \; \bar{\psi} \wedge \gamma ^{ab} \psi =0 , \label{mcadslk} \\
& D_\omega \psi +\frac{1}{4} k^{ab} \wedge \gamma _{ab}\psi +e\; V^a \wedge \gamma_{a} \psi  =0, \label{mcadslpsi}
\end{align}
\end{subequations}
where $D_\omega = d+\omega$ denotes the Lorentz covariant derivative in four dimensions\footnote{In particular, our convention reads: $D_\omega V^{a}= dV^a + \omega^a_{\;b}\wedge V^b$, $D_\omega k^{ab} = d k^{ab} + 2 \omega^a_{\;c} \wedge k^{cb}$, and $D_\omega \psi = d\psi + \frac{1}{4}\omega^{ab}\wedge \gamma_{ab} \psi$.} and $\wedge$ is the wedge product between differential forms. 
Here $\psi$ corresponds to a
Majorana spinor satisfying $\bar{\psi}=\psi^T C$. Note that we have introduced a scale parameter $e=\frac{1}{2l}$, being $l$ the AdS radius. The $1$-form fields of (the dual Maurer-Cartan formulation of) the AdS-Lorentz superalgebra have length dimension $[\omega^{ab}] = L^0$, $[V^a] = L$, $[k^{ab}] = L^0$, and $[\psi]=L^{1/2}$.

We can then define the AdS-Lorentz Lie algebra valued $2$-form supercurvatures as follows (see also \cite{Ipinza:2016con, Ravera:2018zvm, Concha:2015tla}):
\begin{subequations}\label{adslcurvatures}
\begin{align}
\mathcal{R}^{ab} & \equiv d\omega ^{ab}+\omega ^{ac} \wedge \omega _c^{\;b}, \label{rabadsl} \\
R^{a} & \equiv D_\omega V^{a}+k _{\;b}^{a} \wedge V^{b}-\frac{1}{2}\bar{\psi} \wedge \gamma^{a} \psi , \label{raadsl} \\
F^{ab} & \equiv D_\omega k^{ab}+k _{\;c}^{a}\wedge k^{cb}+4 e^{2}\; V^{a} \wedge V^{b}+ e \; \bar{\psi} \wedge \gamma ^{ab} \psi , \label{fabadsl} \\
\Psi  & \equiv D_\omega \psi +\frac{1}{4} k^{ab} \wedge \gamma _{ab}\psi +e\; V^a \wedge \gamma_{a} \psi . \label{psiadsl}
\end{align}
\end{subequations}

Let us now consider the Maurer-Cartan equations associated with the AdS superalgebra $\mathfrak{osp}(4\vert 1)$, which read:
\begin{subequations}\label{mcosp}
\begin{align}
& d\omega ^{ab}+\omega ^{ac} \wedge \omega _c^{\;b} +4 e^{2}\; V^{a} \wedge V^{b}+ e \; \bar{\psi} \wedge \gamma ^{ab} \psi =0, \label{mcospomega} \\
& D_\omega V^{a}-\frac{1}{2}\bar{\psi} \wedge \gamma^{a} \psi =0, \label{mcospv} \\
& D_\omega \psi +e\; V^a \wedge \gamma_{a} \psi  =0. \label{mcosppsi}
\end{align}
\end{subequations}
The corresponding supercurvatures are defined by:
\begin{subequations}\label{ospcurvatures}
\begin{align}
\tilde{R}^{ab} & \equiv d\omega ^{ab}+\omega ^{ac} \wedge \omega _c^{\;b} +4 e^{2}\; V^{a} \wedge V^{b}+ e \; \bar{\psi} \wedge \gamma ^{ab} \psi, \label{rabosp} \\
\tilde{R}^{a} & \equiv D_\omega V^{a}-\frac{1}{2}\bar{\psi} \wedge \gamma^{a} \psi , \label{raosp} \\
\tilde{\Psi} & \equiv D_\omega \psi +e\; V^a \wedge \gamma_{a} \psi ,  \label{psiosp}
\end{align}
\end{subequations}
where we can also write $d\omega ^{ab}+\omega ^{ac} \wedge \omega _c^{\;b} = 	\mathcal{R}^{ab}$. 

Here we have denoted by $\tilde{R}^{ab}$, $\tilde{R}^{a}$, $\tilde{\Psi}$ the $\mathfrak{osp}(4\vert 1)$ supercurvatures in order to avoid confusion with the AdS-Lorentz supercurvatures \eqref{rabadsl}-\eqref{psiadsl} previously introduced.

We can now exploit the freedom of redefining the Lorentz spin connection in $\mathfrak{osp}(4\vert 1)$ by the addition of a new antisymmetric tensor $1$-form field $B^{ab}$ (carrying length dimension zero) as follows:\footnote{On the same lines of what was done in \cite{Andrianopoli:2017itj} in the case of $\mathfrak{osp}(1 \vert 32)$.}
\begin{equation}\label{redef1}
\omega^{ab} \rightarrow \hat{\omega}^{ab} \equiv \omega^{ab} - B^{ab} .
\end{equation}
Let us observe that such a redefinition is always possible and also implies a change of the torsion $2$-form, that is the reason why we will talk about a ``torsion deformation'' of $\mathfrak{osp}(4\vert 1)$.
After having performed the redefinition \eqref{redef1} of the spin connection, if we rename $\hat{\omega}^{ab}$ as $\omega^{ab}$, the Maurer-Cartan equations \eqref{mcospomega}-\eqref{mcosppsi} take the following form:
\begin{subequations}\label{tdmcosp}
\begin{align}
& d\omega ^{ab}+\omega ^{ac} \wedge \omega _c^{\;b} + D_\omega B^{ab} + 
B^a_{\;c}\wedge B^{cb}  +4 e^{2}\; V^{a} \wedge V^{b}+ e \; \bar{\psi} \wedge \gamma ^{ab} \psi =0, \label{tdmcospomega} \\
& D_\omega V^{a} + B^a_{\;b}\wedge V^b -\frac{1}{2}\bar{\psi} \wedge \gamma^{a} \psi =0, \label{tdmcospv} \\
& D_\omega \psi + \frac{1}{4}B^{ab} \wedge \gamma_{ab}\psi +e\; V^a \wedge \gamma_{a} \psi  =0. \label{tdmcosppsi}
\end{align}
\end{subequations}
Now, if we further require, as an extra condition, the Lorentz spin connection $\omega^{ab}$ to satisfy
\begin{equation}\label{minkb}
\mathcal{R}^{ab} = d\omega ^{ab}+\omega ^{ac} \wedge \omega _c^{\;b}  =0 , 
\end{equation}
corresponding to a Minkowski background, then eq. \eqref{tdmcospomega} splits into two equations, namely eq. \eqref{minkb} plus the condition
\begin{equation}\label{kabtd}
D_\omega B^{ab} + B^a_{\;c}\wedge B^{cb} +4 e^{2}\; V^{a} \wedge V^{b}+ e \; \bar{\psi} \wedge \gamma ^{ab} \psi =0 ,
\end{equation}
which defines the Maurer-Cartan equation for the tensor $1$-form field $B^{ab}$.

Observe that the algebra obtained from $\mathfrak{osp}(4\vert 1)$ through the procedure written above is not isomorphic to $\mathfrak{osp}(4\vert 1)$ because of the extra constraint \eqref{minkb}, which implies \eqref{kabtd}, imposed on the Maurer-Cartan equations \eqref{tdmcospomega}-\eqref{tdmcosppsi}.

On the other hand, renaming $B^{ab}$ as $k^{ab}$, we can see that the Maurer-Cartan equations \eqref{minkb}, \eqref{tdmcospv}, \eqref{kabtd}, and \eqref{tdmcosppsi} exactly correspond to those of the AdS-Lorentz superalgebra previously introduced, namely to eqs. \eqref{mcadslomega}-\eqref{mcadslpsi}.
Correspondingly, from \eqref{minkb}, \eqref{tdmcospv}, \eqref{kabtd}, and \eqref{tdmcosppsi} one can also derive the AdS-Lorentz supercurvatures \eqref{rabadsl}-\eqref{psiadsl}.

We can thus conclude that, at the price of introducing the (torsion) field $k^{ab}$ fulfilling \eqref{kabtd}, $\mathfrak{osp}(4\vert 1)$ can be mapped into the AdS-Lorentz superalgebra, where the spin connection $\omega^{ab}$ is identified with the Lorentz connection of a four-dimensional Minkowski space-time with vanishing Lorentz curvature (albeit with a modification of the supertorsion and of the gravitino super field-strength). 
Thus, we can say that the AdS-Lorentz superalgebra can also be viewed as a ``torsion-deformed'' version of $\mathfrak{osp}(4\vert 1)$.\footnote{This was already observed in \cite{Ravera:2018zvm}, but it had not been explicitly derived yet.} Following the prescription we have just described, one could also derive AdS-Lorentz like superalgebras in higher dimensions.

In the sequel, we shall consider the generalized minimal AdS-Lorentz superalgebra \eqref{genadslalg} and carry on an analogous analysis of its relation with $\mathfrak{osp}(4\vert 1)$. 

\subsection{Relation between the generalized AdS-Lorentz superalgebra and $\mathfrak{osp}(4 \vert 1)$}

As we have done in the AdS-Lorentz case, we now describe the generalized AdS-Lorentz superalgebra \eqref{genadslalg} in its dual Maurer-Cartan formulation.
Let us introduce the set of $1$-form fields $\left\{\omega^{ab},V^{a},\tilde{h}^a, \tilde{k}^{ab},k^{ab},\psi, \xi \right\}$ dual to the generators $\left\{J_{ab},P_{a},\tilde{Z}_a, \tilde{Z}_{ab},Z_{ab},Q, \Sigma \right\}$, that is
\begin{equation}
\begin{split}
& \omega^{ab} (J_{cd}) = \delta^{ab}_{cd}, \quad V^a (P_b) = \delta^a_b, \quad \tilde{h}^a (\tilde{Z}_b) = \delta^a_b, \quad \tilde{k}^{ab} (\tilde{Z}_{cd}) = \delta^{ab}_{cd}, \quad k^{ab} (Z_{cd}) = \delta^{ab}_{cd}, \\
& \psi (Q) = \mathfrak{1}, \quad \xi (\Sigma) = \mathfrak{1} .
\end{split}
\end{equation}
Note that the presence of the generators $\tilde{Z}_a$, $\tilde{Z}_{ab}$, $Z_{ab}$, $\Sigma$ implies the introduction of their dual, new, bosonic and fermionic $1$-form fields $\tilde{h}^a$, $\tilde{k}^{ab}$, $k^{ab}$, and $\xi$, respectively. 

The Maurer-Cartan equations describing the generalized AdS-Lorentz superalgebra \eqref{genadslalg} are:
\begin{subequations}\label{mcgen}
\begin{align}
& d\omega ^{ab}+\omega ^{ac} \wedge \omega _c^{\;b} =0, \label{mcgenomega} \\
& D_\omega V^{a}+k _{\;b}^{a} \wedge V^{b} +\tilde{k} _{\;b}^{a} \wedge \tilde{h}^{b}  -\frac{1}{2}\bar{\psi} \wedge \gamma^{a} \psi -\frac{1}{2}\bar{\xi} \wedge \gamma^{a} \xi =0, \label{mcgenv} \\
& D_\omega \tilde{h}^{a}+\tilde{k} _{\;b}^{a} \wedge V^{b} +k _{\;b}^{a} \wedge \tilde{h}^{b}  -\bar{\psi} \wedge \gamma^{a} \xi =0, \label{mcgenh} \\
& D_\omega \tilde{k}^{ab}+2 k _{\;c}^{a}\wedge \tilde{k}^{cb}+8e^{2}\; V^{a} \wedge \tilde{h}^{b} + e \; \bar{\psi} \wedge \gamma ^{ab} \psi + e \; \bar{\xi} \wedge \gamma ^{ab} \xi =0 , \label{mcgenktilde} \\
& D_\omega k^{ab}+\tilde{k} _{\;c}^{a}\wedge \tilde{k}^{cb}+k _{\;c}^{a}\wedge k^{cb}+4 e^{2}\; V^{a} \wedge V^{b}  + 4 e^{2}\; \tilde{h}^{a} \wedge \tilde{h}^{b}+ 2 e \; \bar{\psi} \wedge \gamma ^{ab} \xi =0 , \label{mcgenk} \\
& D_\omega \psi +\frac{1}{4} k^{ab} \wedge \gamma _{ab}\psi +\frac{1}{4} \tilde{k}^{ab} \wedge \gamma _{ab}\xi  + e\; V^a \wedge \gamma_{a} \xi  +e\; \tilde{h}^a \wedge \gamma_{a} \psi =0 , \label{mcgenpsi} \\ 
& D_\omega \xi +\frac{1}{4} k^{ab} \wedge \gamma _{ab} \xi  +\frac{1}{4} \tilde{k}^{ab} \wedge \gamma _{ab}\psi + e\; V^a \wedge \gamma_{a} \psi +e\; \tilde{h}^a \wedge \gamma_{a} \xi  =0, \label{mcgenxi}
\end{align}
\end{subequations}
where both $\psi$ and $\xi$ are Majorana spinors. The $1$-form fields of (the dual Maurer-Cartan formulation of) the generalized AdS-Lorentz superalgebra have length dimension $[\omega^{ab}] = L^0$, $[V^a] = L$, $[\tilde{h}^a] = L$, $[\tilde{k}^{ab}] = L^0$, $[k^{ab}] = L^0$, $[\psi]=L^{1/2}$, and $[\xi]=L^{1/2}$.

We can then define the generalized AdS-Lorentz Lie algebra valued $2$-form supercurvatures as follows (see also \cite{Concha:2015tla}):\footnote{Here, with an abuse of notation, we use the same Greek letters adopted for the case of the AdS-Lorentz superalgebra.}
\begin{subequations}\label{gencurvatures}
\begin{align}
\mathcal{R}^{ab} & \equiv d\omega ^{ab}+\omega ^{ac} \wedge \omega _c^{\;b} , \label{rab} \\
R^a & \equiv D_\omega V^{a}+k _{\;b}^{a} \wedge V^{b} +\tilde{k} _{\;b}^{a} \wedge \tilde{h}^{b}  -\frac{1}{2}\bar{\psi} \wedge \gamma^{a} \psi -\frac{1}{2}\bar{\xi} \wedge \gamma^{a} \xi , \label{ra} \\
\tilde{H}^a & \equiv D_\omega \tilde{h}^{a}+\tilde{k} _{\;b}^{a} \wedge V^{b} +k _{\;b}^{a} \wedge \tilde{h}^{b}  -\bar{\psi} \wedge \gamma^{a} \xi , \label{ha} \\
\tilde{F}^{ab} & \equiv D_\omega \tilde{k}^{ab}+2 k _{\;c}^{a}\wedge \tilde{k}^{cb}+8e^{2}\; V^{a} \wedge \tilde{h}^{b} + e \; \bar{\psi} \wedge \gamma ^{ab} \psi + e \; \bar{\xi} \wedge \gamma ^{ab} \xi  , \label{tildefab} \\
F^{ab} & \equiv D_\omega k^{ab}+\tilde{k} _{\;c}^{a}\wedge \tilde{k}^{cb}+k _{\;c}^{a}\wedge k^{cb}+4 e^{2}\; V^{a} \wedge V^{b}  + 4 e^{2}\; \tilde{h}^{a} \wedge \tilde{h}^{b}+ 2 e \; \bar{\psi} \wedge \gamma ^{ab} \xi , \label{fab} \\
\Psi & \equiv D_\omega \psi +\frac{1}{4} k^{ab} \wedge \gamma _{ab}\psi +\frac{1}{4} \tilde{k}^{ab} \wedge \gamma _{ab}\xi  +e\; V^a \wedge \gamma_{a} \xi  +e\; \tilde{h}^a \wedge \gamma_{a} \psi  , \label{psi} \\ 
\Xi & \equiv D_\omega \xi +\frac{1}{4} k^{ab} \wedge \gamma _{ab}\xi  +\frac{1}{4} \tilde{k}^{ab} \wedge \gamma _{ab}\psi +e\; V^a \wedge \gamma_{a} \psi +e\; \tilde{h}^a \wedge \gamma_{a} \xi  . \label{sigma}
\end{align}
\end{subequations}

Now, considering the Maurer-Cartan equations of $\mathfrak{osp}(4\vert 1)$ given by \eqref{mcospomega}-\eqref{mcosppsi} we observe that, redefining
\begin{equation}\label{redef}
\left\{
\begin{aligned}
& \; \omega^{ab} \rightarrow \hat{\omega}^{ab} \equiv \omega^{ab} - \tilde{B}^{ab} - B^{ab} , \\
& \; V^a \rightarrow \hat{V}^a \equiv V^a - \tilde{B}^a , \\
& \; \psi \rightarrow \hat{\psi} \equiv \psi - \eta ,
\end{aligned}
\right.
\end{equation}
if we then rename $\hat{\omega}^{ab} \Rightarrow \omega^{ab}$, $\hat{V}^a \Rightarrow V^a$, and $\hat{\psi} \Rightarrow \psi$, the Maurer-Cartan equations \eqref{mcospomega}-\eqref{mcosppsi} become:
\begin{subequations}\label{newtdmcosp}
\begin{align}
& d\omega ^{ab} +\omega ^{ac} \wedge \omega _c^{\;b} + D_\omega \tilde{B}^{ab} + D_\omega B^{ab} + \tilde{B}^a_{\;c}\wedge \tilde{B}^{cb} + 2 B^a_{\;c}\wedge \tilde{B}^{cb}+ B^a_{\;c}\wedge B^{cb} + 4 e^{2}\; V^{a} \wedge V^{b} \nonumber \\
& + 8 e^{2}\; V^{a} \wedge \tilde{B}^{b} + 4 e^{2}\; \tilde{B}^{a} \wedge \tilde{B}^{b}+ e \; \bar{\psi} \wedge \gamma ^{ab} \psi + 2 e \; \bar{\psi} \wedge \gamma ^{ab} \eta + e \; \bar{\eta} \wedge \gamma ^{ab} \eta =0, \label{newtdmcospomega} \\
& D_\omega V^{a} + D_\omega \tilde{B}^{a} + B^a_{\;b}\wedge V^b  + B^a_{\;b}\wedge \tilde{B}^b + \tilde{B}^a_{\;b}\wedge V^b + \tilde{B}^a_{\;b}\wedge \tilde{B}^b \nonumber \\
& -\frac{1}{2}\bar{\psi} \wedge \gamma^{a} \psi - \bar{\psi} \wedge \gamma^{a} \eta -\frac{1}{2}\bar{\eta} \wedge \gamma^{a} \eta = 0, \label{newtdmcospv} \\
& D_\omega \psi + D_\omega \eta + \frac{1}{4}B^{ab} \wedge \gamma_{ab}\psi + \frac{1}{4}B^{ab} \wedge \gamma_{ab}\eta + \frac{1}{4}\tilde{B}^{ab} \wedge \gamma_{ab}\psi + \frac{1}{4}\tilde{B}^{ab} \wedge \gamma_{ab}\eta \nonumber \\
& +e\; V^a \wedge \gamma_{a} \psi + e\; V^a \wedge \gamma_{a} \eta + e\; \tilde{B}^a \wedge \gamma_{a} \psi + e\; \tilde{B}^a \wedge \gamma_{a} \eta  =0. \label{newtdmcosppsi}
\end{align}
\end{subequations}
Both $\tilde{B}^{ab}$ and $B^{ab}$ are antisymmetric tensor $1$-forms carrying length dimension zero, $\tilde{B}^a$ is a $1$-form carrying length dimension $1$, and $\eta$ is a spinor $1$-form carrying length dimension $1/2$.

Then, if we further require the Lorentz spin connection $\omega^{ab}$ to satisfy \eqref{minkb} (corresponding to a Minkowski background), together with the following (new) extra conditions:
\begin{subequations}
\begin{align}
& D_\omega \tilde{B}^{a}+\tilde{B} _{\;b}^{a} \wedge V^{b} +B_{\;b}^{a} \wedge \tilde{B}^{b}  -\bar{\psi} \wedge \gamma^{a} \eta =0, \label{cond1} \\
& D_\omega \tilde{B}^{ab}+2 B _{\;c}^{a}\wedge \tilde{B}^{cb}+8e^{2}\; V^{a} \wedge \tilde{B}^{b} + e \; \bar{\psi} \wedge \gamma ^{ab} \psi + e \; \bar{\eta} \wedge \gamma ^{ab} \eta =0 , \label{cond2} \\
& D_\omega B^{ab}+\tilde{B} _{\;c}^{a}\wedge \tilde{B}^{cb}+B _{\;c}^{a}\wedge B^{cb}+4 e^{2}\; V^{a} \wedge V^{b} + 4 e^{2}\; \tilde{B}^{a} \wedge \tilde{B}^{b}+ 2 e \; \bar{\psi} \wedge \gamma ^{ab} \eta =0 , \label{cond3} \\
& D_\omega \eta +\frac{1}{4} B^{ab} \wedge \gamma _{ab}\eta  +\frac{1}{4} \tilde{B}^{ab} \wedge \gamma _{ab}\psi +e\; V^a \wedge \gamma_{a} \psi +e\; \tilde{B}^a \wedge \gamma_{a} \eta =0, \label{cond4}
\end{align}
\end{subequations}
which define the Maurer-Cartan equations for the $1$-form fields $\tilde{B}^a$, $\tilde{B}^{ab}$, $B^{ab}$, and $\eta$, one can easily prove that, after having redefined $\tilde{B}^a \Rightarrow \tilde{h}^a$, $\tilde{B}^{ab} \Rightarrow \tilde{k}^{ab}$, $B^{ab} \Rightarrow k^{ab}$, and $\eta \Rightarrow \xi$, the superalgebra we end up with is exactly the generalized minimal AdS-Lorentz one, with Maurer-Cartan equations given by \eqref{mcgenomega}-\eqref{mcgenxi}. Let us observe that, again, the superalgebra obtained from $\mathfrak{osp}(4\vert 1)$ through the procedure written above (namely, the generalized AdS-Lorentz superalgebra) is not isomorphic to $\mathfrak{osp}(4\vert 1)$, because of the extra constraints \eqref{minkb}, \eqref{cond1}-\eqref{cond4} imposed on the Maurer-Cartan equations \eqref{newtdmcospomega}-\eqref{newtdmcosppsi}. 
One can then define the AdS-Lorentz super field-strengths as given in \eqref{rab}-\eqref{sigma}.

Thus, we can conclude that, at the price of introducing the extra $1$-form fields $\tilde{h}^a$, $\tilde{k}^{ab}$, $k^{ab}$, and $\xi$ (satisfying \eqref{cond1}, \eqref{cond2}, \eqref{cond3}, and \eqref{cond4}, respectively, after having redefined $\tilde{B}^a \Rightarrow \tilde{h}^a$, $\tilde{B}^{ab} \Rightarrow \tilde{k}^{ab}$, $B^{ab} \Rightarrow k^{ab}$, and $\eta \Rightarrow \xi$), $\mathfrak{osp}(4\vert 1)$ can be mapped into the generalized AdS-Lorentz superalgebra, where the spin connection is identified with the Lorentz connection of a Minkowski space-time with vanishing Lorentz curvature (furthermore, we also have a modification of the supertorsion and of the gravitino super field-strength). 
In this sense, the generalized minimal AdS-Lorentz superalgebra can be interpreted as a peculiar ``torsion deformation'' of $\mathfrak{osp}(4\vert 1)$. 

Some comments are in order. Let us first of all observe that the AdS-Lorentz and the generalized minimal AdS-Lorentz superalgebras, which, as we have seen above, correspond to different, peculiar, torsion deformations of $\mathfrak{osp}(4\vert 1)$, can also be both obtained from $\mathfrak{osp}(4\vert 1)$ by performing the so-called $S$-expansion procedure, as it was done in \cite{Concha:2015tla}. In particular, the semigroup leading from $\mathfrak{osp}(4\vert 1)$ to the AdS-Lorentz superalgebra \eqref{adslalg} is the abelian semigroup $S^{(2)}_{\mathcal{M}} = \lbrace \lambda_0 , \lambda_1, \lambda_2 \rbrace$ (according with the notation of \cite{Concha:2015tla}), whose elements obey the multiplication laws
\begin{equation}
\lambda_\alpha \lambda_\beta = 
\left\{
\begin{aligned}
& \; \lambda_{\alpha + \beta} , \quad \text{if} \; \alpha + \beta \leq 2 ,\\
& \; \lambda_{\alpha + \beta -2}, \quad \text{if} \; \alpha + \beta > 2 .
\end{aligned}
\right.
\end{equation}
Similarly, the semigroup leading from $\mathfrak{osp}(4\vert 1)$ to the generalized minimal AdS-Lorentz superalgebra \eqref{genadslalg} is the abelian semigroup $S^{(4)}_{\mathcal{M}} = \lbrace \lambda_0 , \lambda_1, \lambda_2, \lambda_3 , \lambda_4 \rbrace$ (again, according with the notation of \cite{Concha:2015tla}), whose elements obey the following multiplication laws:
\begin{equation}
\lambda_\alpha \lambda_\beta = 
\left\{
\begin{aligned}
& \; \lambda_{\alpha + \beta} , \quad \text{if} \; \alpha + \beta \leq 4 ,\\
& \; \lambda_{\alpha + \beta -4}, \quad \text{if} \; \alpha + \beta > 4 .
\end{aligned}
\right.
\end{equation}
Then, interestingly enough, we can conclude that semigroups of the type $S^{(2n)}_{\mathcal{M}}$ (with $n\geq 1$) can lead from $\mathfrak{osp}(4\vert 1)$ to different torsion deformations of it. We argue that the same should also occur in higher space-time dimensions. 

Let us also observe that, on the other hand, the so-called Maxwell-type superalgebras (commonly related to the AdS-Lorentz type superalgebras through In\"{o}n\"{u}-Wigner contractions), such as those discussed in \cite{Concha:2014tca}, cannot be directly related to $\mathfrak{osp}(4\vert 1)$ by performing a torsion deformation involving a redefinition like \eqref{redef1} or \eqref{redef}.
 
Correspondingly, they can be obtained by performing $S$-expansions of $\mathfrak{osp}(4\vert 1)$ involving semigroups of the type $S^{(2m)}_E$ (with $m\geq 2$), which have different multiplication laws with respect to those of the semigroups $S^{(2n)}_{\mathcal{M}}$ ($n\geq 1$) (see \cite{Concha:2014tca} for details). 
All the above observations could help to shed some light on the relations occurring among the aforementioned different superalgebras and physical theories based on them.

\section{Generalized AdS-Lorentz supergravity in the geometric approach}\label{rheon}

Now, let us briefly recall some of the main features of the rheonomic approach for the description of $\mathcal{N}=1$, $D=4$ pure supergravity (more details can be found in \cite{Andrianopoli:2014aqa, Ipinza:2016con, Ravera:2018zvm}), since this will be useful in the sequel.

In the geometric approach to supergravity \cite{Castellani:1991eu}, the theory is given in terms of $1$-form superfields $\mu^A$ defined on superspace $\mathcal{M}_{4 \vert 4}$. In particular, the bosonic $1$-form $V^a$ and the fermionic $1$-form $\psi^\alpha$ define the supervielbein basis $\lbrace V^a, \psi^\alpha \rbrace$ in superspace.

In this framework, the supersymmetry transformations in space-time are interpreted as diffeomorphisms in the fermionic directions of superspace and they are generated by Lie derivatives with fermionic parameter $\epsilon^\alpha$. Then, the supersymmetry invariance of the theory is fulfilled requiring the Lie derivative of the Lagrangian to vanish for diffeomorphisms in the fermionic directions of superspace, that is to say:
\begin{equation}\label{ell}
\delta_\epsilon \mathcal{L} \equiv \ell _\epsilon \mathcal{L} = \imath_\epsilon d \mathcal{L} + d (\imath_\epsilon \mathcal{L})=0,
\end{equation}
where $\epsilon$ is the fermionic parameter along the tangent vector dual to the gravitino (for simplicity, we have omitted the spinor index $\alpha$), and $\imath$ is
the contraction operator. In particular, we have $\imath_\epsilon (\psi) = \epsilon$ and $\imath_\epsilon (V^a) = 0$.

The contribution $\imath_\epsilon d \mathcal{L}$ in \eqref{ell}, which would be
identically zero in space-time, is non-trivial here, in superspace. On the other hand, the contribution $d (\imath_\epsilon \mathcal{L})$ is a
boundary term and does not affect the bulk result. 
Then, a necessary condition for a supergravity Lagrangian is
\begin{equation}\label{condbulk}
\imath_\epsilon d \mathcal{L} =0 ,
\end{equation}
corresponding to require supersymmetry invariance in the bulk.
Under \eqref{condbulk}, the supersymmetry
transformation of the action simply reduces to 
\begin{equation}
\delta_\epsilon \mathcal{S} = \int_{\mathcal{M}_4} d (\imath_\epsilon \mathcal{L}) = \int_{\partial \mathcal{M}_4} \imath_\epsilon \mathcal{L} .
\end{equation}

When we consider a Minkowski background (or, in general, a space-time with boundary thought as set at infinity), the fields asymptotically vanish, so that 
\begin{equation}\label{condbdy}
\imath_\epsilon \mathcal{L} \vert _{\partial \mathcal{M}_4}=0
\end{equation}
and, consequently, 
\begin{equation}
\delta_\epsilon \mathcal{S}=0 .
\end{equation}
Then, we have that, in this case, eq. \eqref{condbulk} is also a sufficient condition for the supersymmetry invariance of the Lagrangian.
 
On the other hand, when the background space-time presents a non-trivial boundary, the condition \eqref{condbdy} (modulo an exact differential) becomes non-trivial, and it is necessary to check it explicitly to get supersymmetry invariance of the action, requiring a more subtle treatment.

Before analyzing the generalized (minimal) $D=4$ AdS-Lorentz deformed supergravity theory in the presence of a non-trivial boundary of space-time, we will now study the construction of the bulk Lagrangian and the corresponding supersymmetry transformation laws, on the same lines of \cite{Ipinza:2016con}. 
Specifically, we will apply the rheonomic approach to derive the parametrization of the Lorentz-like curvatures involving the extra $1$-form fields $\tilde{h}^a$, $\tilde{k}^{ab}$, $k^{ab}$, and $\xi$ by studying the different sectors of the on-shell Bianchi identities. 
This will also lead to the supersymmetry transformation laws. Subsequently, we will construct a geometric generalized $D=4$ AdS-Lorentz Lagrangian, showing that it can be written in terms of the aforementioned Lorentz-like curvatures (this is an alternative way to introduce a generalized supersymmetric cosmological term, see also \cite{Ipinza:2016con}). After that, we will analyze the supersymmetry invariance of the theory in the presence of a non-trivial space-time boundary.

\subsection{Parametrization of the Lorentz-like curvatures}

Let us consider the following Lorentz-type curvatures defined in superspace:\footnote{Here we use the Greek letters $\tilde{\mathcal{F}}^{ab}$, $\mathcal{F}^{ab}$, $\rho$, and $\sigma$, in order to avoid confusion with the generalized AdS-Lorentz supercurvatures \eqref{tildefab}-\eqref{sigma}.}
\begin{subequations}\label{lorlikecurvatures}
\begin{align}
\mathcal{R}^{ab} & \equiv d\omega ^{ab}+\omega ^{ac} \wedge \omega _c^{\;b} , \label{lrab} \\
R^a & \equiv D_\omega V^{a}+k _{\;b}^{a} \wedge V^{b} +\tilde{k} _{\;b}^{a} \wedge \tilde{h}^{b}  -\frac{1}{2}\bar{\psi} \wedge \gamma^{a} \psi -\frac{1}{2}\bar{\xi} \wedge \gamma^{a} \xi , \label{lra} \\
\tilde{H}^a & \equiv D_\omega \tilde{h}^{a}+\tilde{k} _{\;b}^{a} \wedge V^{b} +k _{\;b}^{a} \wedge \tilde{h}^{b}  -\bar{\psi} \wedge \gamma^{a} \xi , \label{lha} \\
\tilde{\mathcal{F}}^{ab} & \equiv D_\omega \tilde{k}^{ab}+2 k _{\;c}^{a}\wedge \tilde{k}^{cb} , \label{ltildefab} \\
\mathcal{F}^{ab} & \equiv D_\omega k^{ab}+\tilde{k} _{\;c}^{a}\wedge \tilde{k}^{cb}+k _{\;c}^{a}\wedge k^{cb} , \label{lfab} \\
\rho & \equiv D_\omega \psi +\frac{1}{4} k^{ab} \wedge \gamma _{ab}\psi +\frac{1}{4} \tilde{k}^{ab} \wedge \gamma _{ab}\xi , \label{lpsi} \\ 
\sigma & \equiv D_\omega \xi +\frac{1}{4} k^{ab} \wedge \gamma _{ab}\xi  +\frac{1}{4} \tilde{k}^{ab} \wedge \gamma _{ab}\psi  . \label{lsigma}
\end{align}
\end{subequations}
Observe that the supercurvatures \eqref{lrab}-\eqref{lsigma} are actually defined in a superspace that is larger than the ordinary one, whose basis is just given by the supervielbein $\lbrace V^a, \psi \rbrace$. In the sequel, we will ask the parametrization of the curvatures to be well defined in ordinary superspace by exploiting the rheonomic approach.

The supercurvatures \eqref{lrab}-\eqref{lsigma} satisfy the Bianchi identities:
\begin{subequations}\label{bianchi}
\begin{align}
D_\omega \mathcal{R}^{ab} & =0 , \label{b1} \\
D_\omega  R^a & = \mathcal{R}^a_{\;b}\wedge V^b + \mathcal{F}^a_{\;b}\wedge V^b - k^a_{\;b} \wedge R^b + \tilde{\mathcal{F}}^a_{\;b} \wedge \tilde{h}^b - \tilde{k}^a_{\;b}\wedge \tilde{H}^b + \bar{\psi}\wedge \gamma^a \rho + \bar{\xi} \wedge \gamma^a \sigma , \label{b2} \\
D_\omega  \tilde{H}^a & = \mathcal{R}^a_{\;b}\wedge \tilde{h}^b + \tilde{\mathcal{F}}^a_{\;b}\wedge V^b - \tilde{k}^a_{\;b} \wedge R^b + \mathcal{F}^a_{\;b} \wedge \tilde{h}^b  - k^a_{\;b}\wedge \tilde{H}^b + \bar{\xi}\wedge \gamma^a \rho + \bar{\psi} \wedge \gamma^a \sigma , \label{b3} \\
D_\omega  \tilde{\mathcal{F}}^{ab} & = 2 \mathcal{R}^a_{\;c}\wedge \tilde{k}^{cb} + 2 \mathcal{F}^a_{\;c} \wedge \tilde{k}^{cb}+ 2 \tilde{\mathcal{F}}^{a}_{\;c} \wedge k^{cb}  , \label{b4} \\
D_\omega  \mathcal{F}^{ab} & = 2 \mathcal{R}^a_{\;c}\wedge k^{cb} + 2 \tilde{\mathcal{F}}^a_{\;c} \wedge \tilde{k}^{cb}+ 2 \mathcal{F}^{a}_{\;c} \wedge k^{cb} , \label{b5} \\
D_\omega \rho & = \frac{1}{4} \mathcal{R}^{ab} \wedge \gamma_{ab}\psi - \frac{1}{4}\gamma_{ab} \rho \wedge k^{ab} + \frac{1}{4}\mathcal{F}^{ab}\wedge \gamma_{ab}\psi  - \frac{1}{4} \gamma_{ab}\sigma \wedge \tilde{k}^{ab}+ \frac{1}{4}\tilde{\mathcal{F}}^{ab} \wedge \gamma_{ab}\xi  , \label{b6} \\ 
D_\omega \sigma & = \frac{1}{4} \mathcal{R}^{ab} \wedge \gamma_{ab}\xi - \frac{1}{4}\gamma_{ab} \sigma \wedge k^{ab} + \frac{1}{4}\mathcal{F}^{ab}\wedge \gamma_{ab}\xi  - \frac{1}{4} \gamma_{ab}\rho \wedge \tilde{k}^{ab}+ \frac{1}{4}\tilde{\mathcal{F}}^{ab} \wedge \gamma_{ab}\psi . \label{b7}
\end{align}
\end{subequations}
We write the most general ansatz for the Lorentz-type curvatures in the supervielbein basis $\lbrace V^a, \psi \rbrace$ of superspace as follows:
\begin{subequations}\label{ansatzzo}
\begin{align}
\mathcal{R}^{ab} & = \mathcal{R}^{ab}_{\;\;\;\;cd} V^c \wedge V^d + \bar{\Theta}^{ab}_{\;\;\;\;c} \psi \wedge V^c + \alpha e \; \bar{\psi} \wedge \gamma^{ab} \psi , \label{arab} \\
R^a & = R^a_{\;\; bc} V^b \wedge V^c + \bar{\Theta}^a_{\;\;b}\psi \wedge V^b + \beta \bar{\psi} \wedge \gamma^a \psi , \label{ara} \\
\tilde{H}^a & = \tilde{H}^a_{\;\; bc} V^b \wedge V^c + \bar{\Lambda}^a_{\;\;b}\psi \wedge V^b + \gamma \bar{\psi} \wedge \gamma^a \psi , \label{aha} \\
\tilde{\mathcal{F}}^{ab} & = \tilde{\mathcal{F}}^{ab}_{\;\;\;\;cd} V^c \wedge V^d + \bar{\Lambda}^{ab}_{\;\;\;\;c} \psi \wedge V^c + \delta e \; \bar{\psi} \wedge \gamma^{ab} \psi , \label{atildefab} \\
\mathcal{F}^{ab} & = \mathcal{F}^{ab}_{\;\;\;\;cd} V^c \wedge V^d + \bar{\Pi}^{ab}_{\;\;\;\;c} \psi \wedge V^c + \varepsilon e \; \bar{\psi} \wedge \gamma^{ab} \psi , \label{afab} \\
\rho & = \rho_{ab} V^a \wedge V^b + \lambda e \; \gamma_a \psi \wedge V^a + e^{1/2} \; \Omega_{\alpha \beta}\psi^\alpha \wedge \psi^\beta , \label{apsi} \\ 
\sigma & = \sigma_{ab} V^a \wedge V^b + \mu e \; \gamma_a \psi \wedge V^a + e^{1/2} \; \tilde{\Omega}_{\alpha \beta}\psi^\alpha \wedge \psi^\beta  , \label{asigma}
\end{align}
\end{subequations}
where $e$ is the scale parameter (carrying length dimension $-1$) and $\alpha$, $\beta$, $\gamma$, $\delta$, $\varepsilon$, $\lambda$, $\mu$ are coefficients to be determined from the study of the (on-shell) Bianchi identities. Setting $R^a = 0$ (that is called the on-shell condition), we can withdraw some terms appearing in the above ansatz by studying the scaling constraints. On the other hand, the remaining coefficients can be determined from the analysis of the various sectors of the (on-shell) Bianchi identities in superspace \eqref{b1}-\eqref{b7}.

One can then show that the Bianchi identities \eqref{b1}-\eqref{b7} are solved by parametrizing (on-shell) the full set of supercurvatures as follows:
\begin{subequations}\label{paramfs}
\begin{align}
\mathcal{R}^{ab} & = \mathcal{R}^{ab}_{\;\;\;\;cd} V^c \wedge V^d + \bar{\Theta}^{ab}_{\;\;\;\;c} \psi \wedge V^c , \label{prab} \\
R^a & = 0, \label{pra} \\
\tilde{H}^a & = 0 , \label{pha} \\
\tilde{\mathcal{F}}^{ab} & = \tilde{\mathcal{F}}^{ab}_{\;\;\;\;cd} V^c \wedge V^d + \bar{\Lambda}^{ab}_{\;\;\;\;c} \psi \wedge V^c  , \label{ptildefab} \\
\mathcal{F}^{ab} & = \mathcal{F}^{ab}_{\;\;\;\;cd} V^c \wedge V^d + \bar{\Pi}^{ab}_{\;\;\;\;c} \psi \wedge V^c  , \label{pfab} \\
\rho & = \rho_{ab} V^a \wedge V^b , \label{ppsi} \\ 
\sigma & = \sigma_{ab} V^a \wedge V^b  , \label{psigma}
\end{align}
\end{subequations}
with 
\begin{equation}
\begin{split}
\bar{\Theta}^{ab}_{\;\;\;\;c} + \bar{\Pi}^{ab}_{\;\;\;\;c} & = \epsilon ^{abde}\left( \bar{\rho}_{cd}\gamma _{e}\gamma _{5}+\bar{\rho} _{ec}\gamma _{d}\gamma _{5}-\bar{\rho} _{de}\gamma _{c}\gamma _{5}\right) , \\
\bar{\Lambda}^{ab}_{\;\;\;\;c} & = \epsilon ^{abde}\left( \bar{\sigma}_{cd}\gamma _{e}\gamma _{5}+\bar{\sigma} _{ec}\gamma _{d}\gamma _{5}-\bar{\sigma} _{de}\gamma _{c}\gamma
_{5}\right) .
\end{split}
\end{equation}
For reaching this result, we have used the formulas given in Appendix \ref{appa}.
We have thus found the parametrization of the Lorentz-type curvatures \eqref{lrab}-\eqref{lsigma}. This, as we are going to show, also provides us with the supersymmetry transformations laws.

\paragraph{Supersymmetry transformation laws obtained within the geometric approach} 

The parametrizations \eqref{prab}-\eqref{psigma} we have obtained above allow to derive the supersymmetry
transformations in a direct way. 
Indeed, in the geometric framework we have adopted, the transformations
on space-time are given by (see \cite{Castellani:1991eu, Castellani:2018zey} and \cite{Ravera:2018zvm} for details):
\begin{equation}\label{susytrrr}
\delta \mu^A = \left( \nabla \epsilon \right) ^A + \imath_\epsilon R^A,
\end{equation}
for all the superfields $\mu^A$, where the symbol $\nabla$ denotes the gauge covariant derivative and where
\begin{equation}
\epsilon^A \equiv (\epsilon^{ab}, \epsilon^a, \tilde{\varepsilon}^a, \tilde{\varepsilon}^{ab}, \varepsilon^{ab}, \epsilon^\alpha, \varepsilon^\alpha ) .
\end{equation} 

Then, for $\epsilon^{ab} = \epsilon^a = \tilde{\varepsilon}^a = \tilde{\varepsilon}^{ab}= \varepsilon^{ab}=\varepsilon^\alpha =0$, we have (we neglect the spinor index $\alpha$, for simplicity):
\begin{subequations}\label{contr}
\begin{align}
\imath_\epsilon \mathcal{R}^{ab} & = \bar{\Theta}^{ab}_{\;\;\;\;c}\epsilon V^c  , \label{contrrab} \\
\imath_\epsilon R^a & = 0, \label{contrra} \\
\imath_\epsilon \tilde{H}^a & = 0 , \label{contrha} \\
\imath_\epsilon \tilde{\mathcal{F}}^{ab} & = \bar{\Lambda}^{ab}_{\;\;\;\;c} \epsilon V^c  , \label{contrtildefab} \\
\imath_\epsilon \mathcal{F}^{ab} & = \bar{\Pi}^{ab}_{\;\;\;\;c} \epsilon V^c   , \label{contrfab} \\
\imath_\epsilon \rho & = 0 , \label{contrpsi} \\ 
\imath_\epsilon \sigma & = 0 . \label{contrsigma}
\end{align}
\end{subequations}
This provides the following supersymmetry transformation laws for the $1$-form fields:
\begin{subequations}\label{susy}
\begin{align}
\delta_\epsilon \omega^{ab} & = \bar{\Theta}^{ab}_{\;\;\;\;c}\epsilon V^c  , \label{susyom} \\
\delta_\epsilon V^a & = \bar{\epsilon} \gamma^a \psi , \label{susyv} \\
\delta_\epsilon \tilde{h}^a & = \bar{\epsilon} \gamma^a \xi , \label{susyh} \\
\delta_\epsilon \tilde{k}^{ab} & = \bar{\Lambda}^{ab}_{\;\;\;\;c} \epsilon V^c  , \label{susytildek} \\
\delta_\epsilon k^{ab} & = \bar{\Pi}^{ab}_{\;\;\;\;c} \epsilon V^c  , \label{susyk} \\
\delta_\epsilon \psi & = D_\omega \epsilon + \frac{1}{4}\gamma_{ab}\epsilon k^{ab}  , \label{susypsi} \\ 
\delta_\epsilon \xi & = \frac{1}{4} \gamma_{ab} \epsilon \tilde{k}^{ab}  . \label{susyxi}
\end{align}
\end{subequations}
We will now move to the construction of a geometric bulk Lagrangian.

\subsection{Rheonomic construction of the geometric bulk Lagrangian}

We now construct a geometric bulk Lagrangian based on the generalized AdS-Lorentz superalgebra.

The most general ansatz for the aforementioned Lagrangian can be written as follows:
\begin{equation}\label{ansatzl}
\mathcal{L} = \mu^{(4)} + R^A \wedge \mu^{(2)}_A + R^A \wedge R^B \mu^{(0)}_{AB},
\end{equation}
where the upper index ($p$) denotes the degree of the related differential $p$-forms. Here, the $R^A$'s are the generalized AdS-Lorentz Lie algebra valued supercurvatures defined by eqs. \eqref{rab}-\eqref{sigma}, invariant under the rescaling
\begin{equation}
\begin{split}
& \omega^{ab} \rightarrow \omega^{ab}, \quad V^a \rightarrow \omega V^a, \quad \tilde{h}^a \rightarrow \omega \tilde{h}^a, \quad \tilde{k}^{ab} \rightarrow \tilde{k}^{ab}, \\
& k^{ab} \rightarrow k^{ab}, \quad \psi \rightarrow \omega^{1/2} \psi , \quad \xi \rightarrow \omega^{1/2} \xi.
\end{split}
\end{equation}

The Lagrangian must scale with $\omega^2$, being $\omega^2$ the scale-weight of the Einstein-Hilbert term.
Thus, due to scaling constraints reasons (see \cite{Castellani:1991eu}), some of the terms in the ansatz \eqref{ansatzl} disappear.
Besides, since we are now constructing the bulk Lagrangian, we can set $R^A \wedge R^B \mu^{(0)}_{AB}=0$. Nevertheless, these terms will be fundamental for the construction of the boundary contributions needed in order to restore the supersymmetry invariance of the full Lagrangian (understood as bulk plus boundary contributions) in the presence of a non-trivial boundary of space-time. 
Then, applying the scaling and the parity
conservation laws, we are left with the following explicit form for the Lagrangian (written in terms of the generalized AdS-Lorentz $1$-form fields and of the super field-strengths \eqref{rab}-\eqref{sigma}):
\begin{equation}\label{primella}
\begin{split}
\mathcal{L} & = \epsilon_{abcd} \mathcal{R}^{ab} \wedge V^c \wedge V^d + \alpha_1 \epsilon_{abcd} \mathcal{R}^{ab} \wedge V^c \wedge \tilde{h}^d + \alpha_2 \epsilon_{abcd} \mathcal{R}^{ab} \wedge \tilde{h}^c \wedge \tilde{h}^d + \\
& \quad + \alpha_3 \epsilon_{abcd} \tilde{F}^{ab} \wedge V^c \wedge V^d + \alpha_4 \epsilon_{abcd} \tilde{F}^{ab} \wedge V^c \wedge \tilde{h}^d + \alpha_5 \epsilon_{abcd} \tilde{F}^{ab} \wedge \tilde{h}^c \wedge \tilde{h}^d \\
& \quad + \alpha_6 \epsilon_{abcd} F^{ab} \wedge V^c \wedge V^d + \alpha_7 \epsilon_{abcd} F^{ab} \wedge V^c \wedge \tilde{h}^d + \alpha_8 \epsilon_{abcd} F^{ab} \wedge \tilde{h}^c \wedge \tilde{h}^d \\
& \quad + \alpha_9 \bar{\psi} \wedge V^a \gamma_a \gamma_5 \wedge \Psi +  \alpha_{10} \bar{\psi} \wedge \tilde{h}^a \gamma_a \gamma_5 \wedge \Psi +  \alpha_{11} \bar{\psi} \wedge V^a \gamma_a \gamma_5 \wedge \Xi \\
& \quad +  \alpha_{12} \bar{\psi} \wedge \tilde{h}^a \gamma_a \gamma_5 \wedge \Xi  + \alpha_{13} \bar{\xi} \wedge V^a \gamma_a \gamma_5 \wedge \Psi +  \alpha_{14} \bar{\xi} \wedge \tilde{h}^a \gamma_a \gamma_5 \wedge \Psi \\
& \quad + \alpha_{15} \bar{\xi} \wedge V^a \gamma_a \gamma_5 \wedge \Xi +  \alpha_{16} \bar{\xi} \wedge \tilde{h}^a \gamma_a \gamma_5 \wedge \Xi \\
& \quad + e \; \epsilon_{abcd} \bar{\psi} \wedge \gamma^{ab} \psi \wedge ( \beta_1 V^c \wedge V^d + \beta_2 V^c \wedge \tilde{h}^d  + \beta_3  \tilde{h}^c \wedge \tilde{h}^d ) \\
& \quad + e \; \epsilon_{abcd} \bar{\psi} \wedge \gamma^{ab} \xi \wedge ( \beta_4 V^c \wedge V^d + \beta_5 V^c \wedge \tilde{h}^d + \beta_6 \tilde{h}^c \wedge \tilde{h}^d ) \\
& \quad + e \; \epsilon_{abcd}  \bar{\xi} \wedge \gamma^{ab} \xi \wedge (\beta_7 V^c \wedge V^d + \beta_8 V^c \wedge \tilde{h}^d  + \beta_9 \tilde{h}^c \wedge \tilde{h}^d ) \\
& \quad + \beta_{10} e^2 \; \epsilon_{abcd} V^a \wedge V^b \wedge V^c \wedge V^d + \beta_{11} e^2 \; \epsilon_{abcd} V^a \wedge V^b \wedge V^c \wedge \tilde{h}^d \\
& \quad + \beta_{12}e^2 \; \epsilon_{abcd} V^a \wedge V^b \wedge \tilde{h}^c \wedge \tilde{h}^d  + \beta_{13} e^2 \; \epsilon_{abcd} V^a \wedge \tilde{h}^b \wedge \tilde{h}^c \wedge \tilde{h}^d \\
& \quad + \beta_{14}e^2 \; \epsilon_{abcd} \tilde{h}^a \wedge \tilde{h}^b \wedge \tilde{h}^c \wedge \tilde{h}^d ,
\end{split}
\end{equation}
where, in addition, we have consistently set the coefficient of the first term in \eqref{primella} to $1$. The $\alpha_i$'s and the $\beta_j$'s are constant (dimensionless) parameters to be determined by studying the field equations. 

Let us now compute the variation of the Lagrangian with respect to the different fields. Along these calculations, we make use of the formulas given in Appendix \ref{appa}.
The variation of the Lagrangian with respect to the spin connection $\omega^{ab}$ reads
\begin{equation}
\begin{split}
\delta _{\omega} \mathcal{L} & = 2\epsilon_{abcd} \delta \omega^{ab} \wedge  \Bigg[ D_\omega V^c \wedge V^d + \frac{1}{2} \alpha_1 D_\omega V^c \wedge \tilde{h}^d + \frac{1}{2} \alpha_1 D_\omega \tilde{h}^c \wedge V^d + \alpha_2 D_\omega \tilde{h}^c \wedge \tilde{h}^d \\
& \quad + \alpha_3 \tilde{k}^c_{\;f} \wedge V^f \wedge V^d + \alpha_4 \tilde{k}^c_{\;f} \wedge V^f \wedge \tilde{h}^d + \alpha_5 \tilde{k}^c_{\;f} \wedge \tilde{h}^f \wedge \tilde{h}^d + \alpha_6 k^c_{\;f} \wedge V^f \wedge V^d \\
& \quad + \alpha_7 k^c_{\;f} \wedge V^f \wedge \tilde{h}^d + \alpha_8 k^c_{\;f} \wedge \tilde{h}^f \wedge \tilde{h}^d - \frac{1}{8} \alpha_9 \bar{\psi} \wedge \gamma^c \psi \wedge V^d - \frac{1}{8} \alpha_{10} \bar{\psi} \wedge \gamma^c \psi \wedge \tilde{h}^d \\
& \quad - \frac{1}{8} \left(\alpha_{11}+ \alpha_{13} \right) \bar{\psi} \wedge \gamma^c \xi \wedge V^d - \frac{1}{8} \left(\alpha_{12}+ \alpha_{14}\right) \bar{\psi} \wedge \gamma^c \xi \wedge \tilde{h}^d \\
& \quad - \frac{1}{8} \alpha_{15} \bar{\xi} \wedge \gamma^c \xi \wedge V^d - \frac{1}{8} \alpha_{16} \bar{\xi} \wedge \gamma^c \xi \wedge \tilde{h}^d  \Bigg].
\end{split}
\end{equation}
One can then prove that, if 
\begin{equation}
\begin{split}
& \alpha_1=\alpha_4=\alpha_7=2 , \\
& \alpha_2=\alpha_3=\alpha_5=\alpha_6=\alpha_8=1 , \\
& \alpha_9=\alpha_{10}=\alpha_{11}=\alpha_{12}=\alpha_{13}=\alpha_{14}=\alpha_{15}=\alpha_{16}=4 ,
\end{split}
\end{equation}
$\delta_\omega \mathcal{L}=0$ yields the following field equation:
\begin{equation}
\epsilon_{abcd} \left( R^c + \tilde{H}^c \right) \wedge \left( V^d + \tilde{h}^d \right) =0,
\end{equation}
generalizing to $R^c + \tilde{H}^c $ and $V^d + \tilde{h}^d $ the usual equation $\epsilon_{abcd} R^c \wedge V^d =0$ for the supertorsion.
The variation of the Lagrangian with respect to $\tilde{k}^{ab}$ and $k^{ab}$ gives the same result, that is it does not imply any additional on-shell constraint.

Analogously, one can prove that, by setting
\begin{equation}
\begin{split}
& \beta_1 = \beta_3= \beta_7=\beta_9=-1, \\
& \beta_2= \beta_4 = \beta_6=\beta_8 = \beta_{10}= \beta_{14}= -2, \\
& \beta_5 = -4 , \\
& \beta_{11}=\beta_{13} = -8, \\
& \beta_{12}=-12 ,
\end{split}
\end{equation}
the variation of the Lagrangian with respect to the vielbein $V^a$ can be recast into the following form:
\begin{equation}
\begin{split}
\delta_V \mathcal{L} & = [ 2 \epsilon_{abcd} ( \mathcal{R}^{ab} \wedge V^c + \mathcal{R}^{ab} \wedge \tilde{h}^c )  +  2 \epsilon_{abcd}  ( \tilde{F}^{ab} \wedge V^c + \tilde{F}^{ab} \wedge \tilde{h}^c ) \\
& \quad + 2 \epsilon_{abcd} ( F^{ab} \wedge V^c + F^{ab} \wedge \tilde{h}^c ) \\
& \quad +  4 \bar{\psi} \wedge \gamma_d \gamma_5 \Psi +   4 \bar{\psi} \wedge \gamma_d \gamma_5 \Xi +  4 \bar{\xi} \wedge \gamma_d \gamma_5 \Psi +   4 \bar{\xi} \wedge \gamma_d \gamma_5 \Xi ] \wedge \delta V^d .
\end{split}
\end{equation}
Then, $\delta_V \mathcal{L}=0$ leads to the (generalized) equation
\begin{equation}
2 \epsilon_{abcd} ( \mathcal{R}^{ab} + \tilde{F}^{ab} + F^{ab} ) \wedge ( V^c + \tilde{h}^c ) + 4 ( \bar{\psi} + \bar{\xi} \, ) \wedge \gamma_d \gamma_5 (\Psi + \Xi )=0.
\end{equation}
The variation of the Lagrangian with respect to $\tilde{h}^{a}$ yields the same result.
Finally, from the variation of the Lagrangian with respect to the gravitino field $\psi$, we find the (generalized) field equation
\begin{equation}
8 ( V^a + \tilde{h}^a ) \wedge \gamma_a \gamma_5 ( \Psi + \Xi )+ 4 \gamma_a \gamma_5 ( \psi + \xi \, ) \wedge ( R^a + \tilde{H}^a ) =0 .
\end{equation}
The variation with respect to $\xi$ gives the same result.

We have thus completely determined the bulk Lagrangian of the theory, fixing all the coefficients. Interestingly, one can easily prove that the aforementioned geometric bulk Lagrangian can be rewritten in terms of the Lorentz-type curvatures \eqref{lrab}-\eqref{lsigma} as follows:

\begin{equation}\label{bulkL}
\begin{split}
\mathcal{L}_{\text{bulk}} & = \epsilon_{abcd} \mathcal{R}^{ab} \wedge V^c \wedge V^d + 2 \epsilon_{abcd} \mathcal{R}^{ab} \wedge V^c \wedge \tilde{h}^d + \epsilon_{abcd} \mathcal{R}^{ab} \wedge \tilde{h}^c \wedge \tilde{h}^d + \epsilon_{abcd} \tilde{\mathcal{F}}^{ab} \wedge V^c \wedge V^d \\
& \quad + 2 \epsilon_{abcd} \tilde{\mathcal{F}}^{ab} \wedge V^c \wedge \tilde{h}^d + \epsilon_{abcd} \tilde{\mathcal{F}}^{ab} \wedge \tilde{h}^c \wedge \tilde{h}^d + \epsilon_{abcd} \mathcal{F}^{ab} \wedge V^c \wedge V^d \\
& \quad + 2 \epsilon_{abcd} \mathcal{F}^{ab} \wedge V^c \wedge \tilde{h}^d + \epsilon_{abcd} \mathcal{F}^{ab} \wedge \tilde{h}^c \wedge \tilde{h}^d + 4 \bar{\psi} \wedge V^a \gamma_a \gamma_5 \wedge \rho  \\
& \quad + 4 \bar{\psi} \wedge \tilde{h}^a \gamma_a \gamma_5 \wedge \rho +  4 \bar{\psi} \wedge V^a \gamma_a \gamma_5 \wedge \sigma + 4 \bar{\psi} \wedge \tilde{h}^a \gamma_a \gamma_5 \wedge \sigma + 4 \bar{\xi} \wedge V^a \gamma_a \gamma_5 \wedge \rho \\
& \quad + 4 \bar{\xi} \wedge \tilde{h}^a \gamma_a \gamma_5 \wedge \rho +  4 \bar{\xi} \wedge V^a \gamma_a \gamma_5 \wedge \sigma + 4 \bar{\xi} \wedge \tilde{h}^a \gamma_a \gamma_5 \wedge \sigma \\
& \quad + 2 e^2 \; \epsilon_{abcd} V^a \wedge V^b \wedge ( V^c \wedge V^d + 4 V^c \wedge \tilde{h}^d + 6 \tilde{h}^c \wedge \tilde{h}^d ) + 8e^2 \epsilon_{abcd} V^a \wedge \tilde{h}^b \wedge \tilde{h}^c \wedge \tilde{h}^d \\
& \quad + 2e^2 \epsilon_{abcd} \tilde{h}^a \wedge \tilde{h}^b \wedge \tilde{h}^c \wedge \tilde{h}^d \\
& \quad + 2e \epsilon_{abcd}  \bar{\psi} \wedge \gamma^{ab} \psi \wedge ( V^c \wedge V^d + 2 V^c \wedge \tilde{h}^d + \tilde{h}^c \wedge \tilde{h}^d ) \\
& \quad + 4e \epsilon_{abcd} \bar{\psi} \wedge \gamma^{ab} \xi \wedge ( V^c \wedge V^d + 2 V^c \wedge \tilde{h}^d + \tilde{h}^c \wedge \tilde{h}^d ) \\
& \quad + 2e \epsilon_{abcd}  \bar{\xi} \wedge \gamma^{ab} \xi \wedge ( V^c \wedge V^d + 2 V^c \wedge \tilde{h}^d + \tilde{h}^c \wedge \tilde{h}^d ) .
\end{split}
\end{equation}

Notice the presence in \eqref{bulkL} of $e=\frac{1}{2l}$ (being $l$ the AdS radius); the equations of motion of the Lagrangian admit an AdS vacuum solution with cosmological constant (proportional to $e^2$).
Thus, by performing the above procedure, we have introduced a generalized supersymmetric cosmological constant term in a
supergravity theory in an alternative way.

Let us also mention that the Lagrangian \eqref{bulkL} has been written as a first-order Lagrangian, and the field equation for the spin connection $\omega^{ab}$ implies (up to boundary terms) the vanishing, on-shell, of $R^a + \tilde{H}^a$ (defined in eqs. \eqref{lra} and \eqref{lha}, respectively). This is in agreement with the conditions $R^a = 0$ and $\tilde{H}^a = 0$ we have previously imposed in order to find the on-shell supercurvature parametrizations \eqref{prab}-\eqref{psigma} by studying the various sectors of the Bianchi identities.

The space-time Lagrangian \eqref{bulkL} results to be invariant under the supersymmetry transformations \eqref{susyom}-\eqref{susyxi} of the $1$-form fields on space-time, up to boundary terms. As we have already
mentioned, if the space-time background has a non-trivial boundary, we have to check explicitly the condition \eqref{condbdy}.

\section{Supersymmetry invariance of the theory in the presence of a non-trivial boundary of space-time}\label{bdysec}

In the following, we analyze the supersymmetry invariance of the
Lagrangian in the presence of a non-trivial space-time boundary and, in particular, we present the explicit boundary terms required to recover the supersymmetry invariance of the full Lagrangian (given by bulk plus boundary contributions), on the same lines of \cite{Andrianopoli:2014aqa, Ipinza:2016con} (see also \cite{Ravera:2018zvm}). In the calculations presented in this section, we make extensive use of the formulas in four dimensions given in Appendix \ref{appa}.
Thus, let us consider the bulk Lagrangian \eqref{bulkL}.
The supersymmetry invariance in the bulk is satisfied on-shell. Nevertheless, for this theory the boundary invariance of the Lagrangian under supersymmetry is not trivially
satisfied, and the condition \eqref{condbdy} has to be checked in an explicit way in the presence of a non-trivial boundary of space-time.
In fact, we find that, if the fields do not asymptotically vanish at the boundary, we have
\begin{equation}
\imath_\epsilon \mathcal{L}_{\text{bulk}}\vert_{\partial \mathcal{M}} \neq 0 .
\end{equation}

In order to restore the supersymmetry invariance of the theory, it is possible to modify the bulk Lagrangian by adding boundary (i.e. topological) terms, which do not alter the bulk Lagrangian, so
that \eqref{ell} is still fulfilled.
The only possible boundary contributions (that are topological $4$-forms) compatible with parity and Lorentz-like invariance are:
\begin{subequations}\label{ibordi}
\begin{align}
& d \left(\tilde{\omega}^{ab} \wedge \mathcal{N}^{cd} + \tilde{\omega}^{a}_{\;f} \wedge \tilde{\omega}^{fb} \wedge \tilde{\omega}^{cd}  \right) \epsilon_{abcd} = \epsilon_{abcd} \mathcal{N}^{ab} \wedge \mathcal{N}^{cd}, \label{bdy1} \\
& d \left(\bar{\psi} \wedge \gamma_5 \rho + \bar{\xi}\wedge \gamma_5 \sigma + \bar{\psi} \wedge \gamma_5 \sigma + \bar{\xi} \wedge \gamma_5 \rho \right) = \bar{\rho} \wedge \gamma_5 \rho \nonumber \\
& + \bar{\sigma} \wedge \gamma_5 \sigma + 2 \bar{\rho} \wedge \gamma_5 \sigma + \frac{1}{8} \epsilon_{abcd} \mathcal{R}^{ab} \wedge \bar{\psi} \wedge \gamma^{cd} \psi \nonumber \\
& + \frac{1}{8} \epsilon_{abcd} \tilde{\mathcal{F}}^{ab} \wedge \bar{\psi} \wedge \gamma^{cd} \psi + \frac{1}{8} \epsilon_{abcd} \mathcal{F}^{ab} \wedge \bar{\psi} \wedge \gamma^{cd} \psi \nonumber \\
& + \frac{1}{4} \epsilon_{abcd} \mathcal{R}^{ab} \wedge \bar{\psi} \wedge \gamma^{cd} \xi + \frac{1}{4} \epsilon_{abcd} \tilde{\mathcal{F}}^{ab} \wedge \bar{\psi} \wedge \gamma^{cd} \xi \nonumber \\
& + \frac{1}{4} \epsilon_{abcd} \mathcal{F}^{ab} \wedge \bar{\psi} \wedge \gamma^{cd} \xi + \frac{1}{8} \epsilon_{abcd} \mathcal{R}^{ab} \wedge \bar{\xi} \wedge \gamma^{cd} \xi \nonumber \\
& + \frac{1}{8} \epsilon_{abcd} \tilde{\mathcal{F}}^{ab} \wedge \bar{\xi} \wedge \gamma^{cd} \xi + \frac{1}{8} \epsilon_{abcd} \mathcal{F}^{ab} \wedge \bar{\xi} \wedge \gamma^{cd} \xi , \label{bdy2}
\end{align}
\end{subequations}
where we have defined $\tilde{\omega}^{ab} = \omega^{ab} + \tilde{k}^{ab} + k^{ab}$ and $\mathcal{N}^{ab} = \mathcal{R}^{ab} + \tilde{\mathcal{F}}^{ab} + \mathcal{F}^{ab}$.

Then, the boundary terms \eqref{bdy1} and \eqref{bdy2} correspond to the following boundary Lagrangian:
{\small{
\begin{equation}\label{bdylagr}
\begin{split}
\mathcal{L}_{\text{bdy}} & = d \left( H^{(3)} \right)  \\
& = \alpha \epsilon_{abcd} \Bigg( \mathcal{R}^{ab} \wedge \mathcal{R}^{cd} +\tilde{\mathcal{F}}^{ab} \wedge \tilde{\mathcal{F}}^{cd} + \mathcal{F}^{ab} \wedge \mathcal{F}^{cd} + 2 \mathcal{R}^{ab} \wedge \tilde{\mathcal{F}}^{cd} + 2 \mathcal{R}^{ab} \wedge \mathcal{F}^{cd} +  2 \tilde{\mathcal{F}}^{ab} \wedge \mathcal{F}^{cd} \Bigg) \\
& \quad + \beta \Bigg(
\bar{\rho} \wedge \gamma_5 \rho + \bar{\sigma} \wedge \gamma_5 \sigma + 2 \bar{\rho} \wedge \gamma_5 \sigma + \frac{1}{8} \epsilon_{abcd} \mathcal{R}^{ab} \wedge \bar{\psi} \wedge \gamma^{cd} \psi + \frac{1}{8} \epsilon_{abcd} \tilde{\mathcal{F}}^{ab} \wedge \bar{\psi} \wedge \gamma^{cd} \psi \\
& \quad + \frac{1}{8} \epsilon_{abcd} \mathcal{F}^{ab} \wedge \bar{\psi} \wedge \gamma^{cd} \psi + \frac{1}{4} \epsilon_{abcd} \mathcal{R}^{ab} \wedge \bar{\psi} \wedge \gamma^{cd} \xi + \frac{1}{4} \epsilon_{abcd} \tilde{\mathcal{F}}^{ab} \wedge \bar{\psi} \wedge \gamma^{cd} \xi \\
& \quad + \frac{1}{4} \epsilon_{abcd} \mathcal{F}^{ab} \wedge \bar{\psi} \wedge \gamma^{cd} \xi + \frac{1}{8} \epsilon_{abcd} \mathcal{R}^{ab} \wedge \bar{\xi} \wedge \gamma^{cd} \xi + \frac{1}{8} \epsilon_{abcd} \tilde{\mathcal{F}}^{ab} \wedge \bar{\xi} \wedge \gamma^{cd} \xi \\
& \quad + \frac{1}{8} \epsilon_{abcd} \mathcal{F}^{ab} \wedge \bar{\xi} \wedge \gamma^{cd} \xi \Bigg) ,
\end{split}
\end{equation}
}}
where, in fact, 
\begin{equation}
H^{(3)} = \alpha \epsilon_{abcd} \left(\tilde{\omega}^{ab} \wedge \mathcal{N}^{cd} + \tilde{\omega}^{a}_{\;f} \wedge \tilde{\omega}^{fb} \wedge \tilde{\omega}^{cd}  \right) + \beta \left( \bar{\psi} \wedge \gamma_5 \rho + \bar{\xi}\wedge \gamma_5 \sigma + \bar{\psi} \wedge \gamma_5 \sigma + \bar{\xi} \wedge \gamma_5 \rho \right).
\end{equation}
Here, $\alpha$ and $\beta$ are constant parameters. 
Notice that the structure of a supersymmetric Gauss-Bonnet like term appears in \eqref{bdylagr}. 

Then, let us consider the following ``full'' Lagrangian (bulk plus boundary):
{\small{
\begin{equation}\label{full}
\begin{split}
& \mathcal{L}_{\text{full}} = \mathcal{L}_{\text{bulk}} + \mathcal{L}_{\text{bdy}}  \\
& = \epsilon_{abcd} \mathcal{R}^{ab} \wedge V^c \wedge V^d + 2 \epsilon_{abcd} \mathcal{R}^{ab} \wedge V^c \wedge \tilde{h}^d + \epsilon_{abcd} \mathcal{R}^{ab} \wedge \tilde{h}^c \wedge \tilde{h}^d + \epsilon_{abcd} \tilde{\mathcal{F}}^{ab} \wedge V^c \wedge V^d \\
& \quad + 2 \epsilon_{abcd} \tilde{\mathcal{F}}^{ab} \wedge V^c \wedge \tilde{h}^d + \epsilon_{abcd} \tilde{\mathcal{F}}^{ab} \wedge \tilde{h}^c \wedge \tilde{h}^d + \epsilon_{abcd} \mathcal{F}^{ab} \wedge V^c \wedge V^d + 2 \epsilon_{abcd} \mathcal{F}^{ab} \wedge V^c \wedge \tilde{h}^d \\
& \quad + \epsilon_{abcd} \mathcal{F}^{ab} \wedge \tilde{h}^c \wedge \tilde{h}^d + 4 \bar{\psi} \wedge V^a \gamma_a \gamma_5 \wedge \rho +  4 \bar{\psi} \wedge \tilde{h}^a \gamma_a \gamma_5 \wedge \rho +  4 \bar{\psi} \wedge V^a \gamma_a \gamma_5 \wedge \sigma + \\
& \quad + 4 \bar{\psi} \wedge \tilde{h}^a \gamma_a \gamma_5 \wedge \sigma + 4 \bar{\xi} \wedge V^a \gamma_a \gamma_5 \wedge \rho +  4 \bar{\xi} \wedge \tilde{h}^a \gamma_a \gamma_5 \wedge \rho + 4 \bar{\xi} \wedge V^a \gamma_a \gamma_5 \wedge \sigma +  4 \bar{\xi} \wedge \tilde{h}^a \gamma_a \gamma_5 \wedge \sigma \\
& \quad + 2 e^2 \; \epsilon_{abcd} V^a \wedge V^b \wedge V^c \wedge V^d + 8e^2 \epsilon_{abcd} V^a \wedge V^b \wedge V^c \wedge \tilde{h}^d + 12e^2 \epsilon_{abcd} V^a \wedge V^b \wedge \tilde{h}^c \wedge \tilde{h}^d \\
& \quad + 8e^2 \epsilon_{abcd} V^a \wedge \tilde{h}^b \wedge \tilde{h}^c \wedge \tilde{h}^d + 2e^2 \epsilon_{abcd} \tilde{h}^a \wedge \tilde{h}^b \wedge \tilde{h}^c \wedge \tilde{h}^d + 2e \epsilon_{abcd}  \bar{\psi} \wedge \gamma^{ab} \psi \wedge V^c \wedge V^d \\
& \quad + 4e \epsilon_{abcd} \bar{\psi} \wedge \gamma^{ab} \psi \wedge V^c \wedge \tilde{h}^d +  2e \epsilon_{abcd} \bar{\psi} \wedge \gamma^{ab} \psi \wedge \tilde{h}^c \wedge \tilde{h}^d + 4e \epsilon_{abcd} \bar{\psi} \wedge \gamma^{ab} \xi \wedge V^c \wedge V^d \\
& \quad + 8e \epsilon_{abcd}  \bar{\psi} \wedge \gamma^{ab} \xi \wedge V^c \wedge \tilde{h}^d + 4e \epsilon_{abcd} \bar{\psi} \wedge \gamma^{ab} \xi \wedge \tilde{h}^c \wedge \tilde{h}^d + 2e \epsilon_{abcd}  \bar{\xi} \wedge \gamma^{ab} \xi \wedge V^c \wedge V^d \\
& \quad + 4e \epsilon_{abcd}  \bar{\xi} \wedge \gamma^{ab} \xi \wedge V^c \wedge \tilde{h}^d+ 2e \epsilon_{abcd}  \bar{\xi} \wedge \gamma^{ab} \xi \wedge \tilde{h}^c \wedge \tilde{h}^d \\
& \quad + \alpha \epsilon_{abcd} \Bigg( \mathcal{R}^{ab} \wedge \mathcal{R}^{cd} +\tilde{\mathcal{F}}^{ab} \wedge \tilde{\mathcal{F}}^{cd} + \mathcal{F}^{ab} \wedge \mathcal{F}^{cd} + 2 \mathcal{R}^{ab} \wedge \tilde{\mathcal{F}}^{cd} + 2 \mathcal{R}^{ab} \wedge \mathcal{F}^{cd} +  2 \tilde{\mathcal{F}}^{ab} \wedge \mathcal{F}^{cd} \Bigg) \\
& \quad + \beta \Bigg(
\bar{\rho} \wedge \gamma_5 \rho + \bar{\sigma} \wedge \gamma_5 \sigma + 2 \bar{\rho} \wedge \gamma_5 \sigma + \frac{1}{8} \epsilon_{abcd} \mathcal{R}^{ab} \wedge \bar{\psi} \wedge \gamma^{cd} \psi + \frac{1}{8} \epsilon_{abcd} \tilde{\mathcal{F}}^{ab} \wedge \bar{\psi} \wedge \gamma^{cd} \psi \\
& \quad + \frac{1}{8} \epsilon_{abcd} \mathcal{F}^{ab} \wedge \bar{\psi} \wedge \gamma^{cd} \psi + \frac{1}{4} \epsilon_{abcd} \mathcal{R}^{ab} \wedge \bar{\psi} \wedge \gamma^{cd} \xi + \frac{1}{4} \epsilon_{abcd} \tilde{\mathcal{F}}^{ab} \wedge \bar{\psi} \wedge \gamma^{cd} \xi \\
& \quad + \frac{1}{4} \epsilon_{abcd} \mathcal{F}^{ab} \wedge \bar{\psi} \wedge \gamma^{cd} \xi  + \frac{1}{8} \epsilon_{abcd} \mathcal{R}^{ab} \wedge \bar{\xi} \wedge \gamma^{cd} \xi + \frac{1}{8} \epsilon_{abcd} \tilde{\mathcal{F}}^{ab} \wedge \bar{\xi} \wedge \gamma^{cd} \xi \\
& \quad + \frac{1}{8} \epsilon_{abcd} \mathcal{F}^{ab} \wedge \bar{\xi} \wedge \gamma^{cd} \xi \Bigg) .
\end{split}
\end{equation}
}}

Observe that, due to the homogeneous scaling of the Lagrangian, the coefficients $\alpha$ and $\beta$ must be proportional to $e^{-2}$ and $e^{-1}$, respectively (namely they should have length dimension $2$ and $1$, respectively).

Now, the supersymmetry invariance of the full Lagrangian $\mathcal{L}_{\text{full}}$ in \eqref{full}, in the geometric approach, requires
\begin{equation}\label{ellf}
\delta_\epsilon \mathcal{L}_{\text{full}} \equiv \ell _\epsilon \mathcal{L}_{\text{full}} = \imath_\epsilon d \mathcal{L}_{\text{full}} + d (\imath_\epsilon \mathcal{L}_{\text{full}})=0 .
\end{equation}

Since the boundary terms \eqref{bdy1} and \eqref{bdy2} we have introduced so far are total differentials, the condition for supersymmetry in the bulk, that is $\imath_\epsilon d \mathcal{L}_{\text{full}} =0$, is trivially satisfied. 

Then, the supersymmetry invariance of the full Lagrangian $\mathcal{L}_{\text{full}}$ requires just to verify that, for suitable values of $\alpha$ and $\beta$, the condition $\imath_\epsilon  \mathcal{L}_{\text{full}}  =0$ (modulo an exact differential) holds on
the boundary, that is to say $\imath_\epsilon \mathcal{L}_{\text{full}}\vert_{\partial \mathcal{M}}=0$. 

Computing $\imath_\epsilon \mathcal{L}_{\text{full}} $, we get:
\begin{equation}\label{contrLfull}
\begin{split}
\imath_\epsilon  \mathcal{L}_{\text{full}} & = \epsilon_{abcd} \imath_\epsilon \left( \mathcal{R}^{ab} + \tilde{\mathcal{F}}^{ab} + \mathcal{F}^{ab} \right) \wedge V^c \wedge V^d + 2 \epsilon_{abcd} \imath_\epsilon \left( \mathcal{R}^{ab} + \tilde{\mathcal{F}}^{ab} + \mathcal{F}^{ab} \right) \wedge V^c \wedge \tilde{h}^d \\
& \quad + \epsilon_{abcd} \imath_\epsilon \left( \mathcal{R}^{ab} + \tilde{\mathcal{F}}^{ab} + \mathcal{F}^{ab} \right) \wedge \tilde{h}^c \wedge \tilde{h}^d + 4 \bar{\epsilon} \; V^a \wedge \gamma_a \gamma_5 \wedge \rho + 4 \bar{\epsilon} \; \tilde{h}^a \wedge \gamma_a \gamma_5 \wedge \rho \\
& \quad +  4 \bar{\epsilon} \; V^a \wedge \gamma_a \gamma_5 \wedge \sigma + 4 \bar{\epsilon} \; \tilde{h}^a \wedge \gamma_a \gamma_5 \wedge \sigma + 4 \bar{\psi} \wedge V^a \wedge \gamma_a \gamma_5 \imath_\epsilon \left( \rho \right) + 4 \bar{\psi} \wedge V^a \wedge \gamma_a \gamma_5 \imath_\epsilon \left( \sigma \right) \\
& \quad + 4 \bar{\psi} \wedge \tilde{h}^a \wedge \gamma_a \gamma_5 \imath_\epsilon \left( \rho \right) +  4 \bar{\psi} \wedge \tilde{h}^a \wedge \gamma_a \gamma_5 \imath_\epsilon \left( \sigma \right) +  4 \bar{\xi} \wedge V^a \wedge \gamma_a \gamma_5 \imath_\epsilon \left( \rho \right) + \\
& \quad + 4 \bar{\xi} \wedge V^a \wedge \gamma_a \gamma_5 \imath_\epsilon \left( \sigma \right) +  4 \bar{\xi} \wedge \tilde{h}^a \wedge \gamma_a \gamma_5 \imath_\epsilon \left( \rho \right) + 4 \bar{\xi} \wedge \tilde{h}^a \wedge \gamma_a \gamma_5 \imath_\epsilon \left( \sigma \right) \\
& \quad + 4 e \; \epsilon_{abcd} \bar{\epsilon} \gamma^{ab}\psi \wedge V^c \wedge V^d + 8 e \; \epsilon_{abcd} \bar{\epsilon} \gamma^{ab}\psi \wedge V^c \wedge \tilde{h}^d + 4 e \; \epsilon_{abcd} \bar{\epsilon} \gamma^{ab}\psi \wedge \tilde{h}^c \wedge \tilde{h}^d \\
& \quad + 4 e \; \epsilon_{abcd} \bar{\epsilon} \gamma^{ab}\xi \wedge V^c \wedge V^d + 8 e \; \epsilon_{abcd} \bar{\epsilon} \gamma^{ab}\xi \wedge V^c \wedge \tilde{h}^d + 4 e \; \epsilon_{abcd} \bar{\epsilon} \gamma^{ab}\xi \wedge \tilde{h}^c \wedge \tilde{h}^d \\
& \quad + 2 \epsilon_{abcd} \imath_\epsilon \left( \mathcal{R}^{ab} + \tilde{\mathcal{F}}^{ab} + \mathcal{F}^{ab} \right) \wedge \Bigg( \alpha \mathcal{R}^{cd} + \alpha \tilde{\mathcal{F}}^{cd} + \alpha \mathcal{F}^{cd} + \frac{\beta}{16} \bar{\psi} \wedge \gamma^{cd} \psi \\
& \quad + \frac{\beta}{8} \bar{\psi} \wedge \gamma^{cd} \xi + \frac{\beta}{16} \bar{\xi} \wedge \gamma^{cd} \xi  \Bigg) + \frac{\beta}{4} \epsilon_{abcd} \left( \mathcal{R}^{ab} + \tilde{\mathcal{F}}^{ab} + \mathcal{F}^{ab} \right) \wedge \left( \bar{\epsilon} \gamma^{cd} \psi + \bar{\epsilon} \gamma^{cd} \xi \right) \\
& \quad + 2 \beta \imath_\epsilon \left( \bar{\rho} \right) \wedge \gamma_5 \rho + 2 \beta \imath_\epsilon \left( \bar{\sigma} \right) \wedge \gamma_5 \sigma + 2 \beta \imath_\epsilon \left( \bar{\rho} \right) \wedge \gamma_5 \sigma +  2 \beta \imath_\epsilon \left( \bar{\sigma} \right) \wedge \gamma_5 \rho .
\end{split}
\end{equation}
Now, in general, this is not zero, but its projection on the boundary should be (according with what we have previously explained in Section \ref{rheon}). Indeed, in the presence of a non-trivial boundary of space-time, the field equations in superspace for the Lagrangian \eqref{full} acquire non-trivial boundary contributions (besides the contributions to the equations of motion coming from $\mathcal{L}_{\text{bdy}}$, we also have extra contributions from $\mathcal{L}_{\text{bulk}}$, which were neglected in the absence of a boundary, coming from the total differentials originating from partial integration), which lead to the following constraints that are valid on the boundary:
\begin{equation}\label{eqbdy}
\left\{
\begin{aligned}
\left( \mathcal{R}^{ab} + \tilde{\mathcal{F}}^{ab} + \mathcal{F}^{ab} \right)\vert_{\partial \mathcal{M}} = & - \frac{1}{2\alpha} V^a \wedge V^b - \frac{1}{\alpha} V^a \wedge \tilde{h}^b - \frac{1}{2\alpha} \tilde{h}^a \wedge \tilde{h}^b  \\
& - \frac{\beta}{16 \alpha} \bar{\psi} \wedge \gamma^{ab} \psi  - \frac{\beta}{8\alpha} \bar{\psi} \wedge \gamma^{ab} \xi - \frac{\beta}{16 \alpha} \bar{\xi} \wedge \gamma^{ab} \xi  , \\
\left( \rho + \sigma \right)\vert_{\partial \mathcal{M}} = & -\frac{2}{\beta} V^a \wedge \gamma_a \psi -\frac{2}{\beta} V^a \wedge \gamma_a \xi - \frac{2}{\beta} \tilde{h}^a \wedge \gamma_a \psi -\frac{2}{\beta} \tilde{h}^a \wedge \gamma_a \xi .
\end{aligned}
\right.
\end{equation}
We can see that the supercurvatures on the three-dimensional boundary (that is on the contour of the space-time) are not dynamical, rather being fixed to constant values. Notice that these are values in an enlarged anholonomic basis, meaning that the (linear combinations of the) supercurvatures on the boundary are fixed in terms of not only the bosonic and fermionic vielbein ($V^a$ and $\psi$, respectively) but also of the extra bosonic $1$-form field $\tilde{h}^a$ and of the extra fermionic one, $\xi$ (that is in terms of four-dimensional fields). Actually, this should not surprise, since also the Lorentz-like supercurvatures taken as starting point for our geometric construction of the Lagrangian are defined in an enlarged superspace. Nevertheless, as we have previously shown in Section \ref{rheon} by exploiting the rheonomic approach, their parametrization results to be well defined in ordinary superspace.
Thus, in our framework the supersymmetry invariance constrains the boundary values of the supercurvatures (Neumann  boundary conditions) without fixing the superfields themselves on the boundary. 

Then, upon use of \eqref{eqbdy} (and of Fierz identities and gamma matrices formulas reported in Appendix \ref{appa}), after some algebraic manipulation, on the boundary we are left with:
\begin{equation}
\imath_\epsilon \mathcal{L}_{\text{full}} \vert_{\partial \mathcal{M}} = \epsilon_{abcd} \left( 4 e - \frac{\beta}{8 \alpha} - \frac{4}{\beta} \right) \left( \bar{\epsilon} \gamma^{ab} \psi + \bar{\epsilon} \gamma^{ab} \xi  \right) \wedge \left( V^c \wedge V^d + 2 V^c \wedge \tilde{h}^d + \tilde{h}^c \wedge \tilde{h}^d \right).
\end{equation}
Thus, we find that $\imath_\epsilon \mathcal{L}_{\text{full}} \vert_{\partial \mathcal{M}} =0 $ if the following relation between $\alpha$ and $\beta$ holds:
\begin{equation}\label{betaeq}
\frac{\beta}{4\alpha} + \frac{8}{\beta} = 8 e .
\end{equation}
Then, solving eq. \eqref{betaeq} for $\beta$, we obtain:
\begin{equation}\label{square}
\beta = 16 \; e \; \alpha \left(1 \pm \sqrt{1 - \frac{1}{8 \; e^2 \; \alpha}} \right) .
\end{equation}
Now, observe that, by setting the square root in \eqref{square} to zero, which implies
\begin{equation}\label{values}
\alpha = \frac{1}{8 e^2} \quad \Rightarrow \quad \beta = \frac{2}{e},
\end{equation}
we recover the following $2$-form supercurvatures:

\begin{subequations}\label{curvlast}
\begin{align}
N^{ab} & = \mathcal{R}^{ab} + \tilde{\mathcal{F}}^{ab} + \mathcal{F}^{ab} + 8 e^2 \; V^a \wedge \tilde{h}^b + e \; \bar{\psi} \wedge \gamma^{ab} \psi + e \; \bar{\xi} \wedge \gamma^{ab} \xi \nonumber \\
& \quad + 4 e^2 \; V^a \wedge V^b + 4 e^2 \; \tilde{h}^a \wedge \tilde{h}^b + 2 e \; \bar{\psi} \wedge \gamma^{ab} \xi , \label{a} \\
\Omega & = \rho + \sigma + e \; V^a \wedge \gamma_a \xi + e \; \tilde{h}^a \wedge \gamma_a \psi + e \; V^a \wedge \gamma_a \psi + e \; \tilde{h}^a \wedge \gamma_a \xi , \label{b} \\
R^a & = D_\omega V^{a}+k _{\;b}^{a} \wedge V^{b} +\tilde{k} _{\;b}^{a} \wedge \tilde{h}^{b} - \frac{1}{2}\bar{\psi} \wedge \gamma^{a} \psi -\frac{1}{2}\bar{\xi} \wedge \gamma^{a} \xi , \label{c} \\
\tilde{H}^a & = D_\omega \tilde{h}^{a}+\tilde{k} _{\;b}^{a} \wedge V^{b} +k _{\;b}^{a} \wedge \tilde{h}^{b}  -\bar{\psi} \wedge \gamma^{a} \xi . \label{d}
\end{align}
\end{subequations}
Notice that \eqref{a}-\eqref{d} reproduce the generalized AdS-Lorentz supercurvatures, since one can write:
\begin{subequations}\label{curvlastlast}
\begin{align}
N^{ab} & = \mathcal{R}^{ab} + \tilde{F}^{ab} + F^{ab} , \label{aa} \\
\Omega & = \Psi + \Xi , \label{bb} 
\end{align}
\end{subequations}
being $\mathcal{R}^{ab}$, $\tilde{F}^{ab}$, $F^{ab}$, $\Psi$, and $\Xi$ defined in eqs. \eqref{rab}-\eqref{sigma}.

The full Lagrangian \eqref{full}, written in terms of the $2$-form supercurvatures \eqref{aa} and
\eqref{bb}, can be finally recast as a MacDowell-Mansouri like form \cite{MacDowell:1977jt}, that is:
\begin{equation}\label{MMform}
\mathcal{L}_{\text{full}} = \frac{1}{8 e^2} \epsilon_{abcd} N^{ab} \wedge N^{cd} + \frac{2}{e} \bar{\Omega} \wedge \gamma_5 \Omega ,
\end{equation}
whose boundary term, in particular, corresponds to the following supersymmetric Gauss-Bonnet like term (in the sequel, SUSY GB-like term, that is eq. \eqref{bdylagr} in which we have substituted \eqref{values}):

\begin{equation}\label{gbt}
\begin{split}
& \text{SUSY GB-like term} = \frac{1}{8 e^2} \epsilon_{abcd} \Bigg( \mathcal{R}^{ab} \wedge \mathcal{R}^{cd} +\tilde{\mathcal{F}}^{ab} \wedge \tilde{\mathcal{F}}^{cd} + \mathcal{F}^{ab} \wedge \mathcal{F}^{cd} \\
& + 2 \mathcal{R}^{ab} \wedge \tilde{\mathcal{F}}^{cd} + 2 \mathcal{R}^{ab} \wedge \mathcal{F}^{cd} +  2 \tilde{\mathcal{F}}^{ab} \wedge \mathcal{F}^{cd} \Bigg) \\
& + \frac{2}{e} \Bigg(
\bar{\rho} \wedge \gamma_5 \rho + \bar{\sigma} \wedge \gamma_5 \sigma + 2 \bar{\rho} \wedge \gamma_5 \sigma + \frac{1}{8} \epsilon_{abcd} \mathcal{R}^{ab} \wedge \bar{\psi} \wedge \gamma^{cd} \psi + \frac{1}{8} \epsilon_{abcd} \tilde{\mathcal{F}}^{ab} \wedge \bar{\psi} \wedge \gamma^{cd} \psi \\
& + \frac{1}{8} \epsilon_{abcd} \mathcal{F}^{ab} \wedge \bar{\psi} \wedge \gamma^{cd} \psi  + \frac{1}{4} \epsilon_{abcd} \mathcal{R}^{ab} \wedge \bar{\psi} \wedge \gamma^{cd} \xi + \frac{1}{4} \epsilon_{abcd} \tilde{\mathcal{F}}^{ab} \wedge \bar{\psi} \wedge \gamma^{cd} \xi \\
& + \frac{1}{4} \epsilon_{abcd} \mathcal{F}^{ab} \wedge \bar{\psi} \wedge \gamma^{cd} \xi + \frac{1}{8} \epsilon_{abcd} \mathcal{R}^{ab} \wedge \bar{\xi} \wedge \gamma^{cd} \xi + \frac{1}{8} \epsilon_{abcd} \tilde{\mathcal{F}}^{ab} \wedge \bar{\xi} \wedge \gamma^{cd} \xi \\
& + \frac{1}{8} \epsilon_{abcd} \mathcal{F}^{ab} \wedge \bar{\xi} \wedge \gamma^{cd} \xi \Bigg) .
\end{split}
\end{equation}

Let us observe that considering the square root in \eqref{square} as different from zero would cause other boundary terms appearing in the MacDowell-Mansouri like Lagrangian. Indeed, defining $f^2 = 1 - \frac{1}{8 \; e^2 \; \alpha}$ and considering $f \neq 0$ in \eqref{square} ($\beta \neq 0 $ $\Rightarrow$ $f \neq -1$), we end up with the following extra contributions:
\begin{equation}\label{nuovi}
\begin{split}
& - \frac{f^2}{8 e^2 (f^2-1)} d \left(\tilde{\omega}^{ab} \wedge \mathcal{N}^{cd} + \tilde{\omega}^{a}_{\;f} \wedge \tilde{\omega}^{fb} \wedge \tilde{\omega}^{cd}  \right) \epsilon_{abcd} \\
& +16 e \alpha f d \left(\bar{\psi} \wedge \gamma_5 \rho + \bar{\xi} \wedge \gamma_5 \sigma+ \bar{\psi} \wedge \gamma_5 \sigma + \bar{\xi} \wedge \gamma_5 \rho \right)  
\end{split}
\end{equation}
(recall that we defined $\tilde{\omega}^{ab} = \omega^{ab} + \tilde{k}^{ab} + k^{ab}$ and $\mathcal{N}^{ab} = \mathcal{R}^{ab} + \tilde{\mathcal{F}}^{ab} + \mathcal{F}^{ab}$).
These terms break the off-shell generalized AdS-Lorentz structure of the theory.
However, the first term in \eqref{nuovi} is incompatible with the invariance of the Lagrangian under diffeomorphisms in the bosonic directions of superspace; on the other hand, considering the second term in \eqref{nuovi} and using the value of $\rho + \sigma$ at the boundary, given in \eqref{eqbdy}, we can easily prove that this term vanishes on-shell. Thus, in view of the fact that the closure of the generalized minimal AdS-Lorentz superalgebra only holds on-shell for a superymmetric theory (in the absence of auxiliary fields), this extra contribution does not play a significant role as far as supersymmetry is concerned.

We have thus shown that the Gauss-Bonnet like term given in \eqref{gbt} allows to recover the supersymmetry invariance of the (on-shell) generalized AdS-Lorentz deformed supergravity theory in the presence of a non-trivial boundary of space-time. 

Observe that, in terms of the newly defined supercurvatures \eqref{a} and \eqref{b}, the boundary conditions on the super field-strengths \eqref{eqbdy} take the following simple form: $N^{ab}\vert_{\partial \mathcal{M}} = 0$ and $\Omega \vert_{\partial \mathcal{M}} = 0$. This means, in particular, that the linear combinations $\mathcal{R}^{ab} + \tilde{F}^{ab} + F^{ab}$ and $\Psi + \Xi$ vanish at the boundary.

\section{Comments and possible developments}\label{conclusions}

In this paper, driven by the results of \cite{Andrianopoli:2014aqa} and \cite{Ipinza:2016con}, we have presented the explicit geometric construction
of the $D = 4$ generalized (minimal) AdS-Lorentz deformed supergravity bulk Lagragian (based on the generalized minimal AdS-Lorentz superalgebra of \cite{Concha:2015tla}).
In particular, we have introduced in an alternative way a generalized supersymmetric cosmological
term and we have studied the supersymmetry invariance of the Lagrangian in the presence of a non-trivial boundary of space-time, finding that the supersymmetric
extension of a Gauss-Bonnet like term is required in order to restore the supersymmetry invariance of the full Lagrangian (understood as bulk plus boundary terms).
In this way, we have also further investigated on the study performed in \cite{Concha:2015tla} in the context of AdS-Lorentz superalgebras and generalized supersymmetric cosmological constant terms in $\mathcal{N}=1$ supergravity.


The presence of the $1$-form fields $\tilde{k}^{ab}$, $k^{ab}$, and $\xi$ in the boundary could be useful in the context of the AdS/CFT correspondence. 
In particular, as it was shown in \cite{Miskovic:2009bm}, the introduction of a topological boundary in a four-dimensional bosonic action is equivalent to the holographic renormalization procedure in the AdS/CFT context. 
Then, we conjecture that the presence of $\tilde{k}^{ab}$, $k^{ab}$, and $\xi$ in the boundary of our theory, allowing to recover the supersymmetry invariance in the geometric approach, could also allow to regularize the (deformed) supergravity action in the holographic renormalization context.
Furthermore, it would also be interesting to discuss our construction in the context of the recent works \cite{Anastasiou:2017xjr, Anastasiou:2018rla}.

In this work, we have also observed that both the AdS-Lorentz and the generalized minimal AdS-Lorentz superalgebras can be viewed as peculiar torsion deformations of $\mathfrak{osp}(4\vert 1)$. This is intriguing, since, on the other hand, the same superalgebras can be obtained through $S$-expansion from $\mathfrak{osp}(4\vert 1)$ by using semigroups of the type $S^{(2n)}_{\mathcal{M}}$, with $n \geq 1$ ($S^{(2)}_{\mathcal{M}}$ and $S^{(4)}_{\mathcal{M}}$, respectively, see \cite{Concha:2015tla} for details). 
Then, our results could be useful to shed some light on the properties and physical role of these semigroups, also in higher-dimensional cases. 
Moreover, the form of the MacDowell-Mansouri like action obtained in \cite{Concha:2015tla} by considering the generalized minimal AdS-Lorentz superalgebra coincides with the one in \eqref{MMform} of our paper, obtained by adopting a geometric approach. We argue that all the superalgebras which can be obtained through $S$-expansion from $\mathfrak{osp}(4\vert 1)$ by using semigroups of the type $S^{(2n)}_{\mathcal{M}}$ ($n \geq 1$) can be viewed as particular torsion deformations of $\mathfrak{osp}(4\vert 1)$, in the sense intended in this paper, and that they can consequently lead to MacDowell-Mansouri like actions involving supersymmetric extension of Gauss-Bonnet like terms allowing the supersymmetry invariance of the full Lagrangians (bulk plus boundary contributions) in the presence of a non-trivial boundary of space-time. 

Then, it would also be compelling to analyze differences and analogies (from a geometric point of view) between the case we have discussed in the present work and the case of the super-Maxwell algebras, such as the minimal super-Maxwell algebra of \cite{Concha:2014tca} (called $s\mathcal{M}_4$ in the same paper). 
In particular, in \cite{Concha:2014tca} the authors obtained the minimal $D=4$ supergravity action plus boundary terms from the minimal Maxwell superalgebra $s\mathcal{M}_4$ applying the $S$-expansion procedure to $\mathfrak{osp}(4\vert 1)$. Let us observe, as a first hint towards this possible future study, that the action they ended up with can be also viewed as an In\"{o}n\"{u}-Wigner contraction of \eqref{MMform}, and, on the other hand, it cannot be written as a sum of quadratic terms in the super field-strengths considered in \cite{Concha:2014tca}.

Another future analysis could consist in investigating the possible relations among the extra $1$-form fields appearing in the generalized minimal AdS-Lorentz superalgebra (and also those of the super-Maxwell type algebras) and the extra $1$-forms appearing in the hidden superalgebras underlying supergravity theories in higher dimensions \cite{DAuria:1982uck, Andrianopoli:2016osu, Andrianopoli:2017itj, Ravera:2018vra} (see also \cite{Ravera:2018zvm}). 
This analysis could also shed some light on the conjectured relations \cite{Andrianopoli:2016osu, Andrianopoli:2017itj} between the aforementioned hidden superalgebras and the framework of Exceptional Field Theory (see \cite{Hohm:2013pua, Hohm:2013uia, Hohm:2014qga} and references therein). Some work is in progress on this topic.
Moreover, it would be also interesting to discuss AdS-Lorentz (and also super-Maxwell) deformed supergravity theories in the context of gauged supergravities, exploiting the powerful formalism of the embedding tensor \cite{Trigiante:2016mnt}.

Finally, one could also carry on a further analysis in order to shed some light on the boundary theory produced in our geometric approach. In this context, let us stress that in our framework the supersymmetry invariance constrains the boundary values of the supercurvatures (Neumann  boundary conditions), without fixing, however, the superfields themselves on the boundary. The boundary conditions obtained within our approach are still written in terms of four-dimensional fields and give the values of the curvatures on the three-dimensional boundary, that is on the contour of the four-dimensional space-time, while in order to discuss the theory living on the boundary (in the spirit of the AdS/CFT correspondence, where the supergravity fields act as sources for the CFT operators) one should set the boundary at infinity (that is at $r \rightarrow \infty$, being $r$ the radial coordinate) and study the asymptotic limit $r \rightarrow \infty$ of the $D=3$ equations on the boundary. The explicit three-dimensional description of the equations we have found in $D=4$ would depend on the general symmetry properties of the theory on the boundary, which can be obtained as an effective theory on an asymptotic boundary placed at $r \rightarrow \infty$. One should properly choose the boundary behavior of the $D=4$ fields which relates them to the $D = 3$ ones and perform the asymptotic limit $r \rightarrow \infty$.\footnote{See for example the analysis recently presented in \cite{Andrianopoli:2018ymh}, where the authors found unexpected intriguing relations between $\mathcal{N} = 2$, $D=4$ supergravity and a three-dimensional theory describing the properties of graphene.} 
Since such a study goes beyond the aim of our current paper and would require a lot of work and further calculations, we leave it as a future development. 
Nevertheless, we can conjecture that in the scenario of our paper, where, in particular, we have the presence of a non-trivial boundary of space-time (meaning that the boundary is not thought as set at infinity and thus the fields do not asymptotically vanish) and of extra bosonic and fermionic $1$-form fields appearing both in the bulk and in the boundary contributions to the $D=4$ Lagrangian, the related three-dimensional boundary theory could feature some generalization of deformed locally AdS$_3$ geometries, due to the presence of extra $D=4$ fields from the very beginning.

\section*{Acknowledgements}

We are grateful to Alberto Santambrogio and Dietmar Klemm for the support. L.R. also acknowledges interesting discussions with Laura Andrianopoli, Riccardo D'Auria, and Mario Trigiante.

\vspace{0.2cm}

\noindent
Version accepted for publication in The European Physical Journal Plus (EPJP), Eur. Phys. J. Plus (2018) \textbf{133}: 514. The final publication is available at Springer via \\ https://doi.org/10.1140/epjp/i2018-12335-0.

\appendix

\section{Useful formulas in four dimensions}\label{appa}

The gamma matrices in $D=4$ space-time dimensions are defined through $
\lbrace{ \gamma_a , \gamma_b }\rbrace = - 2 \eta_{ab}$, where $\eta_{ab}$ is the Minkowski metric (we adopt the convention $\eta_{ab} \equiv (-1,1,1,1)$).
They satisfy the algebraic relations:
\begin{equation}\label{gammaform}
\begin{split}
& [\gamma_a , \gamma_b] = 2 \gamma_{ab} , \quad \gamma_5  = - \gamma_0 \gamma_1 \gamma_2 \gamma_3 , \quad \gamma^2_5 = -1 , \\
& \lbrace{ \gamma_5 , \gamma_a }\rbrace  = [\gamma_5 , \gamma_{ab}] =0, \quad
\gamma_{ab} \gamma_5 = -\frac{1}{2}\epsilon_{abcd} \gamma^{cd}, \\
& \gamma_a \gamma_b  = \gamma_{ab}-\eta_{ab}, \quad \gamma^{ab} \gamma_{cd} = \epsilon^{ab}_{\;\;\;cd} \gamma_5 - 4  \delta^{[a}_{\;\;[c} \gamma^{b]}_{\;\;d]}- 2 \delta^{ab}_{cd}, \\
& \gamma^{ab} \gamma^c = 2 \gamma^{[a}\delta^{b]}_c - \epsilon^{abcd} \gamma_5 \gamma_d , \\
& \gamma^c \gamma^{ab} = - 2 \gamma^{[a}\delta^{b]}_c - \epsilon^{abcd}\gamma_5 \gamma_d , \quad \gamma_m \gamma^{ab} \gamma^m  = 0, \\
& \gamma_{ab} \gamma_m \gamma^{ab} =0, \quad \gamma_{ab}\gamma_{cd}\gamma^{ab} = 4 \gamma_{cd}, \quad  \gamma_m \gamma^a \gamma^m  = -2 \gamma^a .
\end{split}
\end{equation}
Furthermore, we have:
\begin{equation}\label{gammasymm}
\begin{split}
& (C \gamma_a)^T = C \gamma_a , \quad (C \gamma_{ab})^T  = C \gamma_{ab} , \\
& (C \gamma_5)^T = - C \gamma_5 , \quad (C \gamma_5 \gamma_a)^T  = - C \gamma_5 \gamma_a  ,
\end{split}
\end{equation}
where $C$ is the charge conjugation matrix ($C^T = - C$). We are dealing with Majorana spinors, fulfilling $\bar{\psi} = \psi^T C$. The following identities hold:
\begin{equation}\label{symm}
\begin{split}
& \bar{\psi} \wedge \xi = (-1)^{pq}\bar{\xi} \wedge \psi , \\
& \bar{\psi} \wedge S \xi = - (-1)^{pq} \bar{\xi} \wedge S \psi , \\
& \bar{\psi} \wedge A \xi  = (-1)^{pq}\bar{\xi} \wedge A \psi 
\end{split}
\end{equation}
for the $p$-form $\psi$ and $q$-form $\xi$, being $S$ and $A$ symmetric and antisymmetric matrices, respectively. 
Finally, we can write the following Fierz identities in four dimensions:
\begin{subequations}\label{fierzids}
\begin{align}
& \psi \wedge \bar{\psi} = \frac{1}{2} \gamma_a \bar{\psi} \wedge \gamma^a \psi - \frac{1}{8} \gamma_{ab} \bar{\psi} \wedge \gamma^{ab} \psi , \label{f1} \\
& \gamma_a \psi \wedge \bar{\psi} \wedge \gamma^a \psi = 0, \label{f2} \\
& \gamma_{ab} \psi \wedge \bar{\psi} \wedge \gamma^{ab} \psi = 0, \label{f3} \\
& \gamma_{ab} \psi \wedge \bar{\psi} \wedge \gamma^a \psi = \psi \wedge \bar{\psi} \wedge \gamma_b \psi . \label{f4}
\end{align}
\end{subequations}

\end{document}